\documentclass[aps,twocolumn,superscriptaddress,longbibliography,numerical]{revtex4-1}

\usepackage{amsmath}
\usepackage{epsfig}
\usepackage{graphicx}
\usepackage{bm}
\usepackage{amssymb}
\usepackage{slashed}
\usepackage{hyperref}
\hypersetup{
     colorlinks   = true,
     citecolor    = blue,
     urlcolor = blue,
     linkcolor = blue
}

\begin{document}
% ==============================================================================

\title{Disorder-induced exceptional points and nodal lines in Dirac superconductors}

\author{Alexander~A.~Zyuzin}
\affiliation{Department of Applied Physics, Aalto University, P.~O.~Box 15100, FI-00076 AALTO, Finland}
\affiliation{Ioffe Physical--Technical Institute,~194021 St.~Petersburg, Russia}

\author{Pascal Simon}
\affiliation{Laboratoire de Physique des Solides, CNRS, Univ. Paris-Sud, University Paris-Saclay, 91405 Orsay Cedex, France}

% ==============================================================================
%\date{\today}

\begin{abstract}
We consider the effect of disorder on the spectrum of quasiparticles in the point-node and nodal-line superconductors. 
Due to the anisotropic dispersion of quasiparticles disorder scattering  may render the Hamiltonian describing these excitations non-Hermitian.  
Depending on the dimensionality of the system, we show that the nodes in the spectrum are replaced by Fermi arcs or Fermi areas bounded by exceptional points or exceptional lines, respectively. 
These features are illustrated by first considering a model of a proximity-induced superconductor in an anisotropic two-dimensional (2D) Dirac semimetal, 
where a Fermi arc in the gap  bounded by  exceptional points can  be realized. 
We next show that the interplay between disorder and supercurrents can give rise to a 2D Fermi surface bounded by exceptional lines  in  three-dimensional (3D) nodal superconductors.
\end{abstract}
\maketitle

\section{Introduction}
The physics of non-Hermitian systems and of exceptional points typically arises in open quantum systems with energy gain  and loss.  Their complex spectra can present exceptional points \cite{Kato}  which occur when the line-widths of two neighboring resonance frequencies coalesce under some fine-tuning of the system parameters \cite{Bender, Berry,Heiss}.
Such physics has been realized experimentally  in various  open quantum systems where gain and loss can be introduced
in a controllable manner, for example in circuits of resonators \cite{Stehmann, Luo}, in photonic systems  \cite{Longhi, Regensburger, Malzard, flach, Invisible_PT, Reflectionless_PT, Whisp_gallery_PT}, or in cold atomic gas \cite{Duan_first}.
Signatures of exceptional points and loops were recently observed in optical waveguides \cite{Soljacic2015, Soljacic-science2018, Cerjan_exp}.

In contrast, non-Hermiticity can also naturally emerge in a disordered and/or interacting closed system if we are interested in single-particle excitations. In the presence of disorder, a single-particle excitation of a given momentum acquires a finite life time. One can thus associate a non-Hermitian Hamiltonian in order to describe these single-particle excitations \cite{Mudry, Kozii, Zyuzin_second,Papaj_Fu, Yoshida, Moors}.

Because topology has completely reshaped our understanding of electronic band structure in solids,
 there has been in the past years a growing interest in analyzing the topological
properties of non-Hermitian Hamiltonians \cite{Esaki2011,Liang2013,Lee2016, Nori2017,Duan_first,Kozii,Shen-Fu,Papaj_Fu,Ueda2018, Zhong, LoicHerviou, New_1, New_3}.
The  topologically-nontrivial electronic bands are
often characterized by the presence of Dirac 
touching points or line nodes.  A natural question is how these band structures are modified for  non-Hermitian Hamiltonians.
In the simplest non-Hermitian extension of a 2D Dirac semimetal, the quasiparticle
conduction and valence bands  are connected by a bulk Fermi arc that ends
at two topological exceptional points instead of a touching Dirac point \cite{Kozii}. At these singularities, the non-Hermitian Hamiltonian becomes defective. 

This extends to a 3D point node semimetal where the two bands, by adding a non-Hermitian term to the Hamiltonian, can stick on a surface bounded by a  one-dimensional (1D) loop of exceptional points \cite{Duan_first, Zyuzin_second, Cerjan_third}. In 3D nodal-line semimetals, the nodal lines can be split into two exceptional lines, connected by a Fermi ribbon \cite{Zyuzin_second, Johan_Emil_forth, Yang_fith, Wang_sixth,  Moors, Okugawa, Budich,New_2}.
In these cases, the real part of the spectrum with momenta lying within the 1D arc (the 2D  area) bounded by the exceptional points (or loops) is zero.  
More generally, it has been shown that  $D$-dimensional non-Hermitian systems can support up to 
$(D-1)$-dimensional exceptional surfaces thanks to parity-time or parity-particle-hole symmetries \cite{Budich, Okugawa}.

Recently, the interplay of topology and non-Hermiticity was extended on  one-dimensional superconductors with a particular emphasis on the difference between Andreev and Majorana bound sates \cite{Pikulin_jetplet,Pikulin_prb,Aguado_1, Ueda_SC, Aguado_2}.
Here, we focus on the band dispersion  of the quasiparticles excitations in 2D and 3D nodal superconductors in the presence of weak disorder. 
As for the semimetallic case, we find that the complex self-energy correction to the Green function of 
quasiparticles due to disorder gives rise to  non-Hermitian Bogoliubov-de Gennes (BdG) Hamiltonians describing  quasiparticle bands with exceptional points and lines.

To be more specific, we first consider the proximity induced superconductivity in a 2D Dirac semimetal with a tilted Dirac cone. We find a phase transition as a function of the Dirac cone anisotropy,
between the gapped state and the state where the superconducting gap closes in momentum space at the Fermi arc bounded by two exceptional points. We then analyze the effect of weak disorder on 
3D nodal-ring and point-node superconductors in the presence of the supercurrent flow. We find that the nodal structure in the spectrum of quasiparticles extends into the flat-band regions bounded by the 
exceptional lines. 

Our paper is structured as follows. Section II contains a short reminder of disorder averaging within the Born approximation. In Sec. III, we consider the effects of disorder on 2D Dirac semimetals in proximity of a superconductor. In Sec IV, we deal with the effects of disorder scattering in 3D nodal superconductors. Finally, in Sec. V we summarize and discuss our results.

\section{Disorder scattering}
Let us first recall the effect of the electron elastic scattering on the random scalar potential disorder on the spectrum of nodal superconductors.
One of the prominent effects of the disorder on the unusual superconductivity is to smear the nodes in the superconducting gap in momentum space, which can be described by the imaginary part of the self-energy correction to the disorder averaged Green function, Refs. \cite{Gorkov, Mineev_book}.
We consider the short-range impurity scattering potential of the form $V(\mathbf{r}) = u_0 \Sigma_a\delta(\mathbf{r}-\mathbf{r}_a)$, where the sum is over the all impurities coordinates $\mathbf{r}_a$.
The energy and the decay-time of the quasiparticles are defined by the real and imaginary parts of the poles of the disorder-averaged Green function in the superconducting state, respectively. 

To find these poles, we will be utilizing the self-consistent equation for the quasiparticle's retarded self-energy $\Sigma(E)$ (a $4\times4$ matrix written in Nambu representation to include spin and particle-hole degrees of freedom), which within Born approximation is given by
\begin{equation}\label{Sigma_main}
\Sigma(E) = \gamma \int \frac{d^dk}{(2\pi)^d}\tau_z [E - H(\mathbf{k}) -\Sigma(E)]^{-1}\tau_z,
\end{equation}
where $\gamma = u_0^2 n_0$ defines the strength of the scattering potential, in which $n_0$ is the impurity concentration, $H(\mathbf{k})$ is the BCS Hamiltonian of the system in the Nambu representation (the explicit expressions for the Hamiltonian of 2D Dirac semimetals in proximity of a superconductor and of 3D nodal superconductors will be specified further in the text), $\tau_{x,y,z}$ are the Pauli matrices in the Nambu space,
and $d=2,3$ is the dimensionality of the sample. 

In this case of the impurity potential the self-energy is momentum independent. A different potential shape would only render the integration over momentum in Eq. 1 more complicated, but not change the following results of the paper qualitatively. We also neglect the interference of waves scattered on different impurities.

The poles of the disorder averaged Green function can be found from the equation 
\begin{equation}\label{eq:poles}
\mathrm{det}[E - H(\mathbf{k}) -\Sigma(E)]=0.
\end{equation} 

Generally, the self-energy might render $H(\mathbf{k}) - \Sigma(E)$ to be a non-Hermitian matrix 
with complex eigenvalues and eventually  exceptional points in the frequency-momentum space.  
This implies that the solutions of Eq. (\ref{eq:poles}), which correspond to quasiparticle excitations of momentum $k$, acquire now a finite energy width.
In the disordered case, the nodal touching points in the spectrum of quasiparticles may transform into exceptional points, at which the non-Hermitian matrix is defective and hence the spectrum has a singularity.
Let us now discuss the conditions needed for the realization of exceptional points in the spectrum of nodal superconductors for several representative cases.

\section{Interplay between disorder and proximity-induced superconductivity in a 2D Dirac semimetal}
We start with a  model of a s-wave superconductor placed in contact with a 2D Dirac semimetal with linear dispersion around the band-touching points \cite{Fu_Kane, Beenakker_RevModPhys}. 
The proximity effect can lead to a gap in the density of states.
The linearized Hamiltonian around the single Dirac point (we will neglect disorder scattering between the Dirac valleys) in the presence of the proximity induced superconducting gap is given by
%\begin{equation}\label{2D_gap}
%H(\mathbf{k}) = \kappa v k_x + (v[\boldsymbol{\sigma}\times\mathbf{k}] \cdot \hat{z}-\mu) \tau_z + \Delta\tau_x,
%\end{equation}
\begin{equation}\label{2D_gap}
H(\mathbf{k}) = \kappa v k_x + [v(\sigma_xk_y-\sigma_yk_x)-\mu] \tau_z + \Delta\tau_x,
\end{equation}
where $\sigma_{x,y}$ are the Pauli matrices acting in the spin space, $\mu$ is the chemical potential, $\kappa$ is a dimensionless parameter related to the inclination of the Dirac cone ($\kappa=0$ means no tilt),  $v$ is the Fermi velocity in absence of tilt, 
and $\Delta$ is the proximity induced superconducting gap which is considered to be a positive real constant here.
We set $\hbar =1$ throughout our calculations.

\begin{figure}[t]
\begin{tabular}{cc}
\includegraphics[width=4cm]{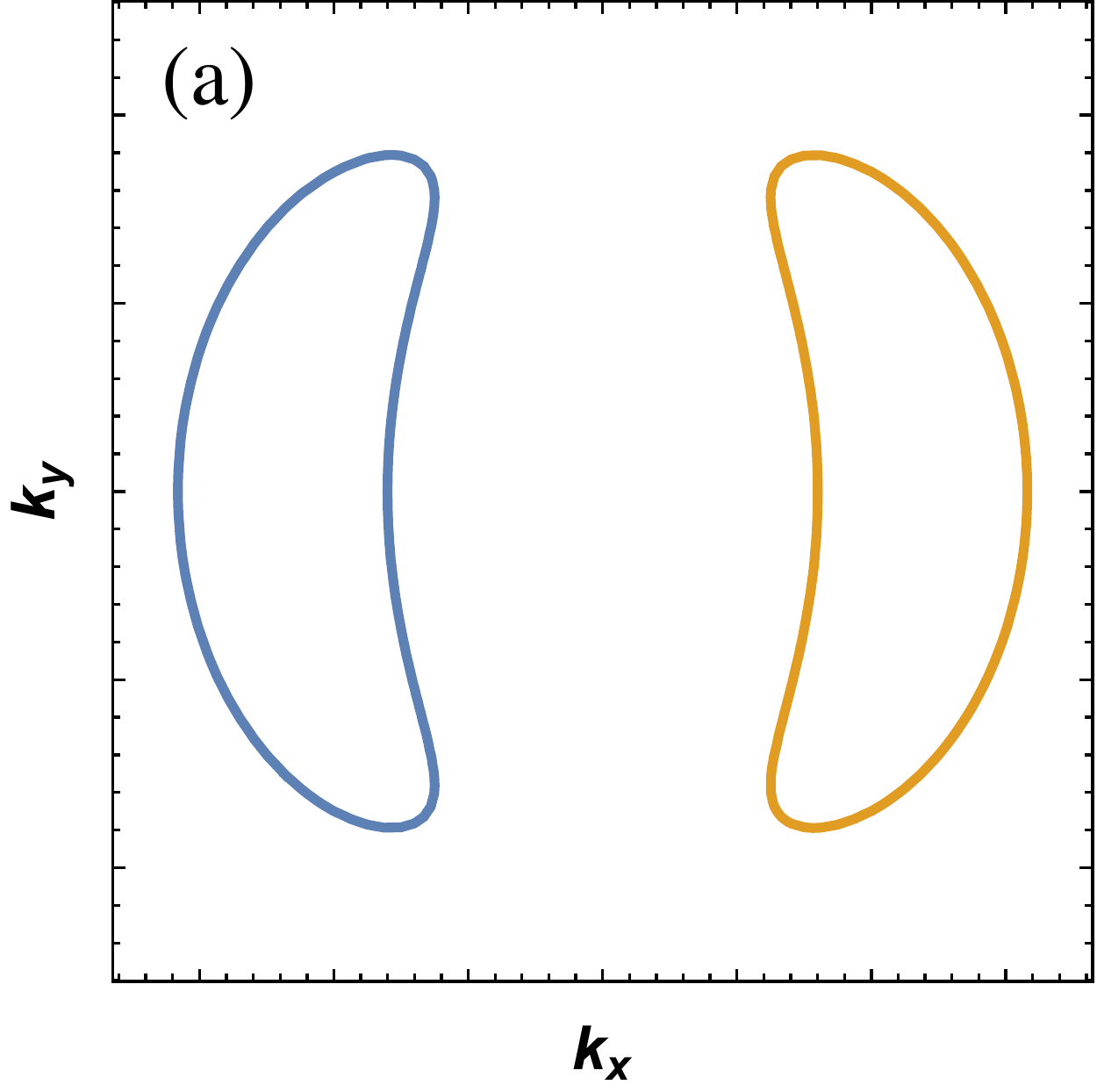}\includegraphics[width=4cm]{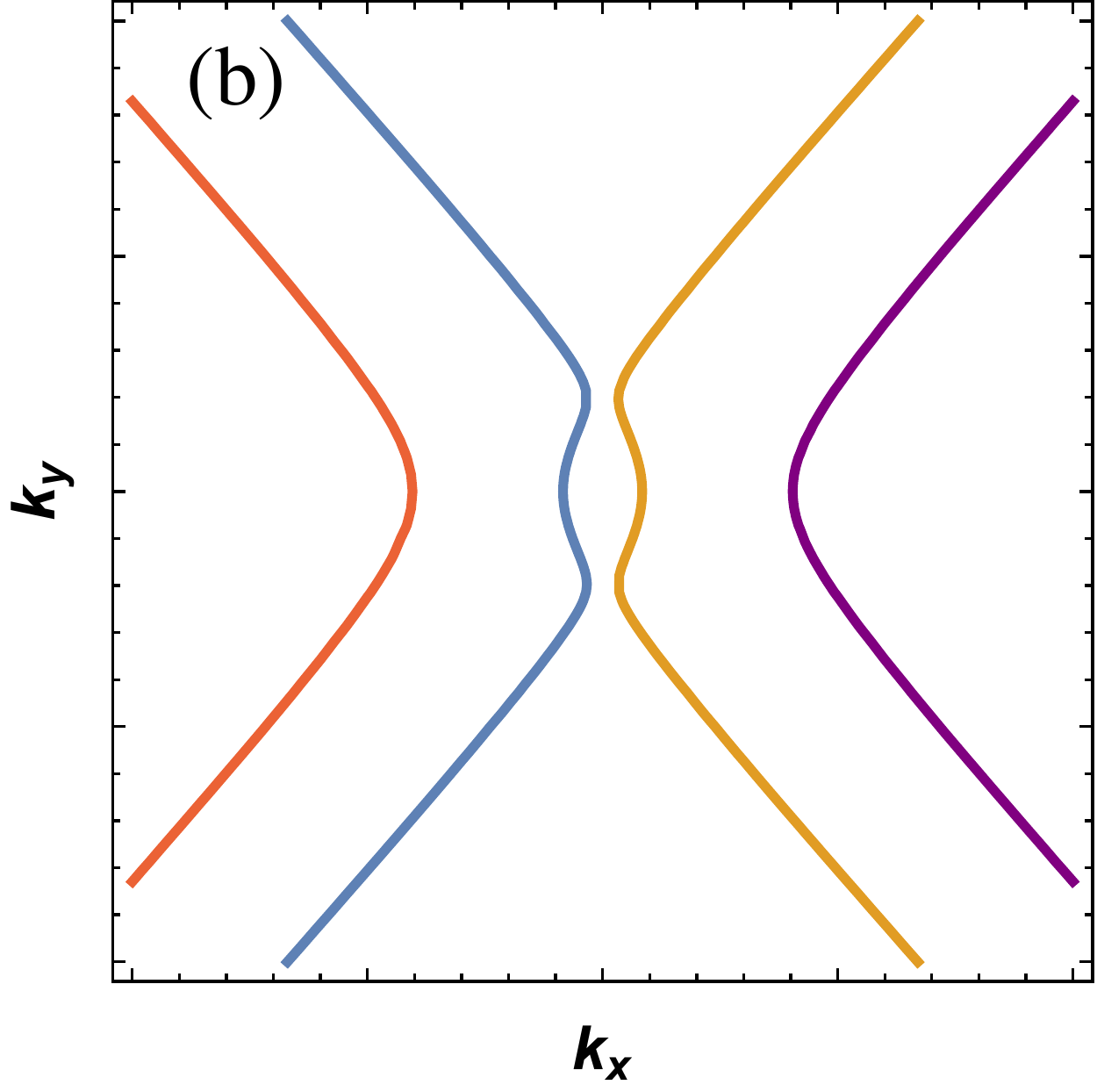}\\
\includegraphics[width=4cm]{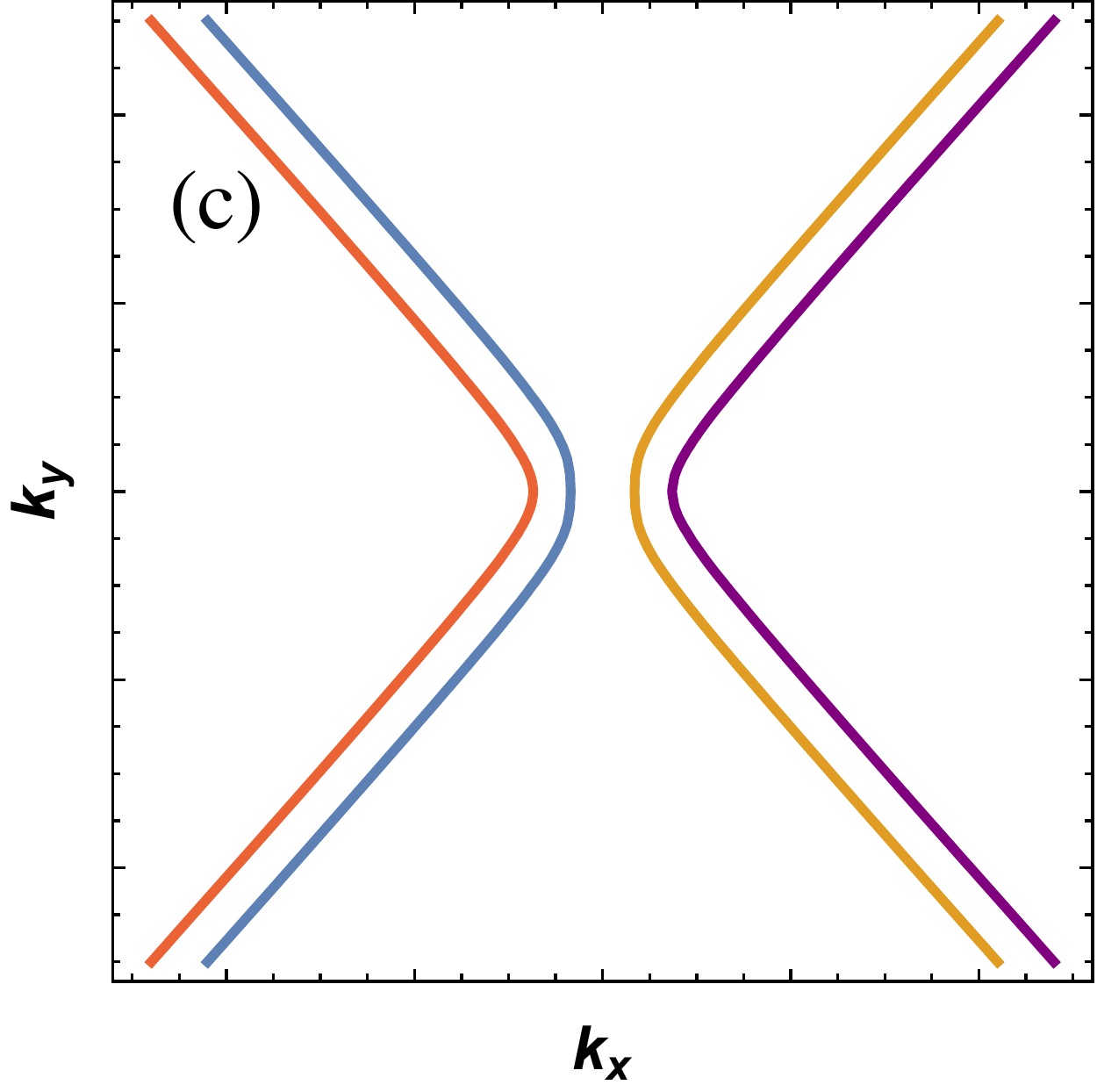} \includegraphics[width=4cm]{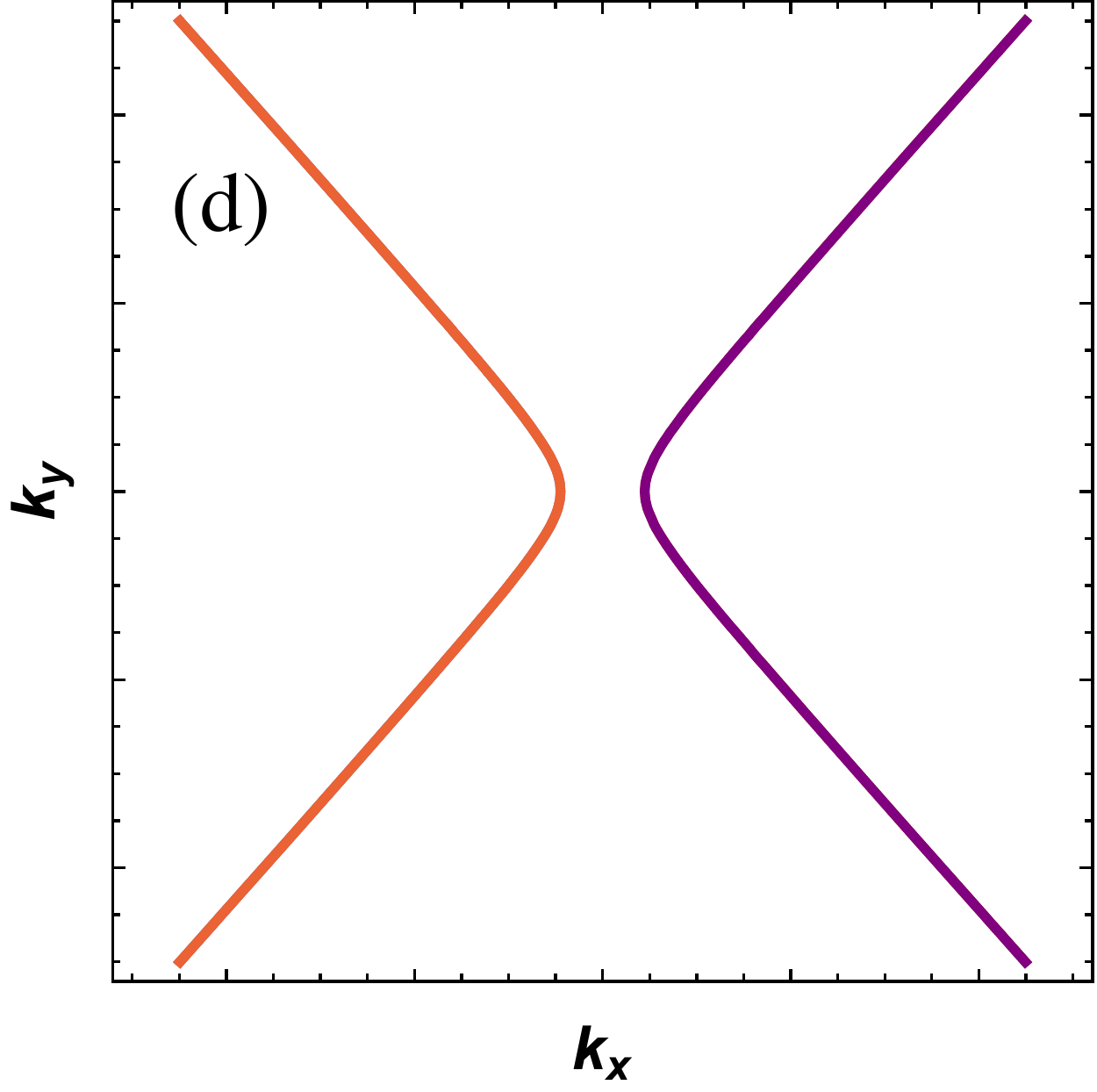}
\end{tabular}
\caption{The lines of zeros in the spectrum of quasiparticles of 2D Dirac semimetal with tilted Dirac cone and proximity-induced superconductivity. The plots show the zero solutions for the spectrum of quasiparticles, given explicitly by Eq. (\ref{EE}) in the text. Panels (a) and (b) show the solutions at two tilt parameters $\kappa = 0.4$ and $\kappa = 1.5$, respectively, for a fixed $\Delta$ and the chemical potential $\mu =4 \Delta$. The nodal-lines exist at small tilt parameter $\Delta/\mu < |\kappa| < 1$. Panels (c) and (d) show the nodal lines in the spectrum at $\mu = 0.5 \Delta$ and $\mu = 0$, respectively, at fixed $\Delta$ and $\kappa =1.5$.} 
\label{fig_new}
\end{figure}

The spectrum of quasiparticles is given by 
\begin{equation}\label{EE}
E_{\pm,s}(\mathbf{k}) = \kappa v k_x \pm \sqrt{(vk-s\mu)^2+\Delta^2},
\end{equation}
 where $s=\pm$. The solution of $E_{\pm,s}(\mathbf{k})=0$ as a function of model parameters is shown in Fig. \ref{fig_new}. At $\mu=0$, the spectrum is gapped provided $|\kappa|<1$ and hosts nodal lines in the opposite case, $|\kappa| > 1$ \cite{Goerbig_2d_tilt}. The nodal lines are determined 
by the equation 
\begin{equation}
(\kappa^2-1)v^2k_x^2 = v^2k_y^2 + \Delta^2.
\end{equation}
It is worth noting that the linearized Hamiltonian, for which the nodal lines are hyperbolas, is applicable only for momenta below some tilt dependent cutoff $k_0 \gg \Delta/v\sqrt{\kappa^2-1}$ determining the width of the electron and hole pockets.
When the chemical potential is placed in the conduction band (here, it is enough to consider $s=1$), 
the transition between the gapped and nodal lines state takes place at much smaller values of the tilt parameter 
$|\kappa| = \frac{\Delta}{\mu}\ll 1$; see Fig. \ref{fig_new}(a).  We will focus below on the case of low doping $\mu \rightarrow 0$, shown in Fig. \ref{fig_new}(d), where the exceptional nodal lines can be realized.

\subsection{Normal case $\Delta=0$}
Before dealing with the superconducting case, let us first recall results about the normal case $\Delta=0$ \cite{Kozii,Papaj_Fu,Zhao_Liu}.
At $\kappa=0$, the Dirac point is smeared due to weak scalar disorder \cite{Shon_Ando}.
To proceed, we consider two limiting cases of weak $|\kappa| < 1$ and strong $|\kappa| > 1$ inclinations of the Dirac cone. 
We search for the self-energy correction to the disorder averaged Green function of the quasiparticles in the presence of the proximity induced superconducting gap by substituting the expression for the Hamiltonian Eq. (\ref{2D_gap}) into the Eq. (\ref{Sigma_main}).

We recover, as it was shown in Refs. \cite{Papaj_Fu,Zhao_Liu}, that in the presence of a finite but small tilt such that $|\kappa|< 1$,
the self-energy in the normal case at zero frequency acquires a nontrivial matrix structure $\Sigma(0) = -i  \left(1 + \kappa \sigma_y \tau_z \right) \Gamma_1$ (note that the self-energy in particle-hole space is described by a $\tau_z$ matrix), where
\begin{eqnarray}\label{Gamma_1}
\Gamma_1= 2 \Lambda \frac{\sqrt{1-\kappa^2}}{1+\kappa^2} \exp\left[-\frac{2\pi v^2}{\gamma}\frac{(1-\kappa^2)^{3/2}}{1+\kappa^2}\right].~~~
\end{eqnarray}
Here, $\Lambda$ is the energy cut-off corresponding to the separation between the Dirac point and the closest bulk band (see Appendix \ref{sec:self-energy_A1} for more details). 
The quasiparticles spectrum is found by solving Eq. (\ref{eq:poles}) which  now becomes complex valued, $E_{\pm}(\mathbf{k})=\kappa v k_x - i \Gamma_1 \pm \sqrt{v^2k_y^2+(vk_x + i\kappa \Gamma_1)^2}$.
Hence, the dispersion of quasiparticles contains a line segment $k_y \in [- |\kappa|\Gamma_1, |\kappa|\Gamma_1]$ at $k_x=0$  bounded by two exceptional points. Such a segment in the quasiparticles spectrum is characterized by a vanishing real part,  the so-called bulk Fermi arc \cite{Kozii}. This means that the decay rate of a quasiparticle has a strong spatial anisotropy. 

In the limit of strong tilt $|\kappa| > 1$, introducing a cutoff $\propto k_0$ for the width of the electron-hole pockets in momentum space, we obtain in the limit $|\kappa| k_0 \gg |E|$,
\begin{equation}\label{Sigma_2}
\Sigma(E) = -i \Gamma_2 \left(1+ \frac{\sigma_y}{\kappa}\tau_z \right) + O(E/|\kappa| k_0),
\end{equation}
where $ \Gamma_2 = \frac{\gamma}{2\pi v}\frac{ |\kappa| k_0 }{ \sqrt{\kappa^2-1}}$  the scattering rate, in which the density of states is determined by $k_0(\kappa)$. 
Importantly, the self-energy contains a frequency independent imaginary contribution, which also results in an 
unusual bulk Fermi arc in the spectrum of quasiparticles \cite{Kozii}.

It is worth noting that the linearized model introduced in Eq. (\ref{2D_gap}) can not be applied in the limit $|\kappa| \rightarrow 1$ due to
unavoidable higher-order corrections in momentum, which are not taken into account here.

\begin{figure}[t]
\begin{tabular}{cc}
\includegraphics[width=4cm]{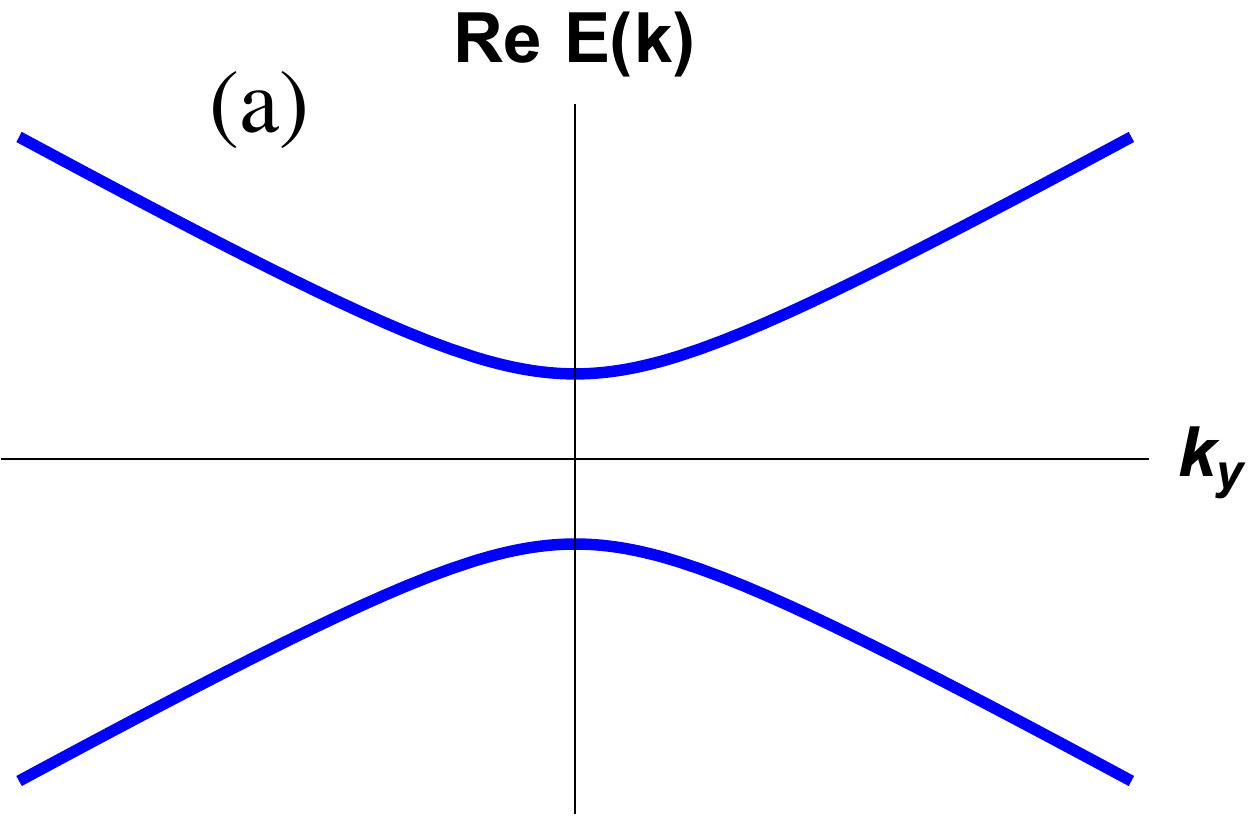} \includegraphics[width=4cm]{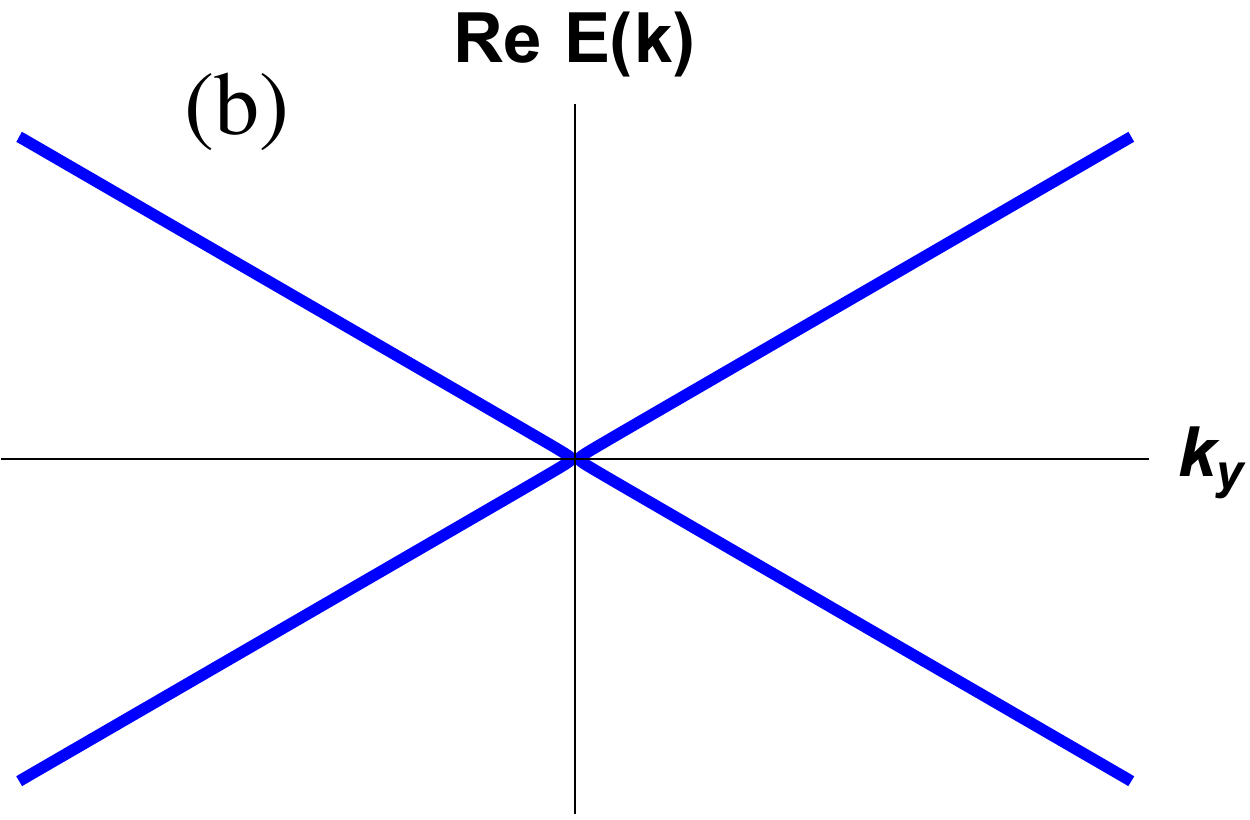}\\
\includegraphics[width=4cm]{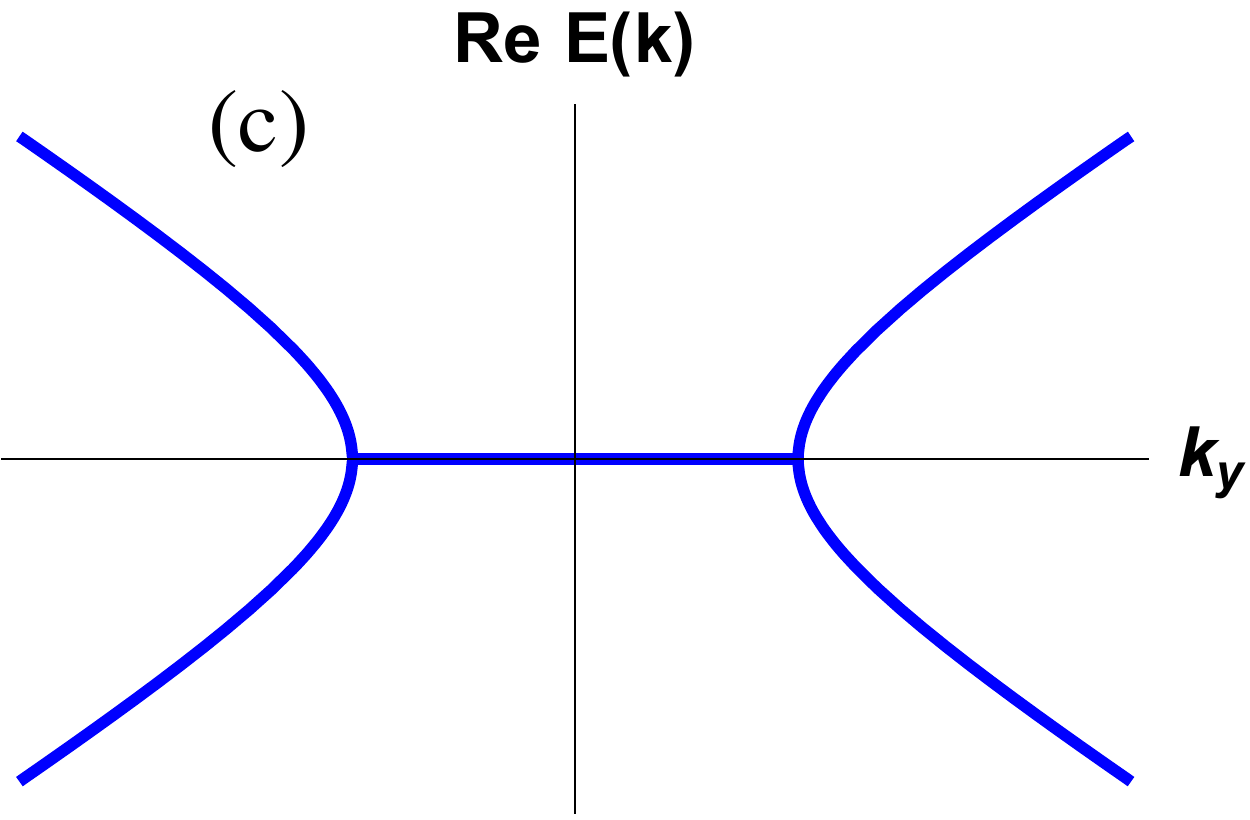} \includegraphics[width=4cm]{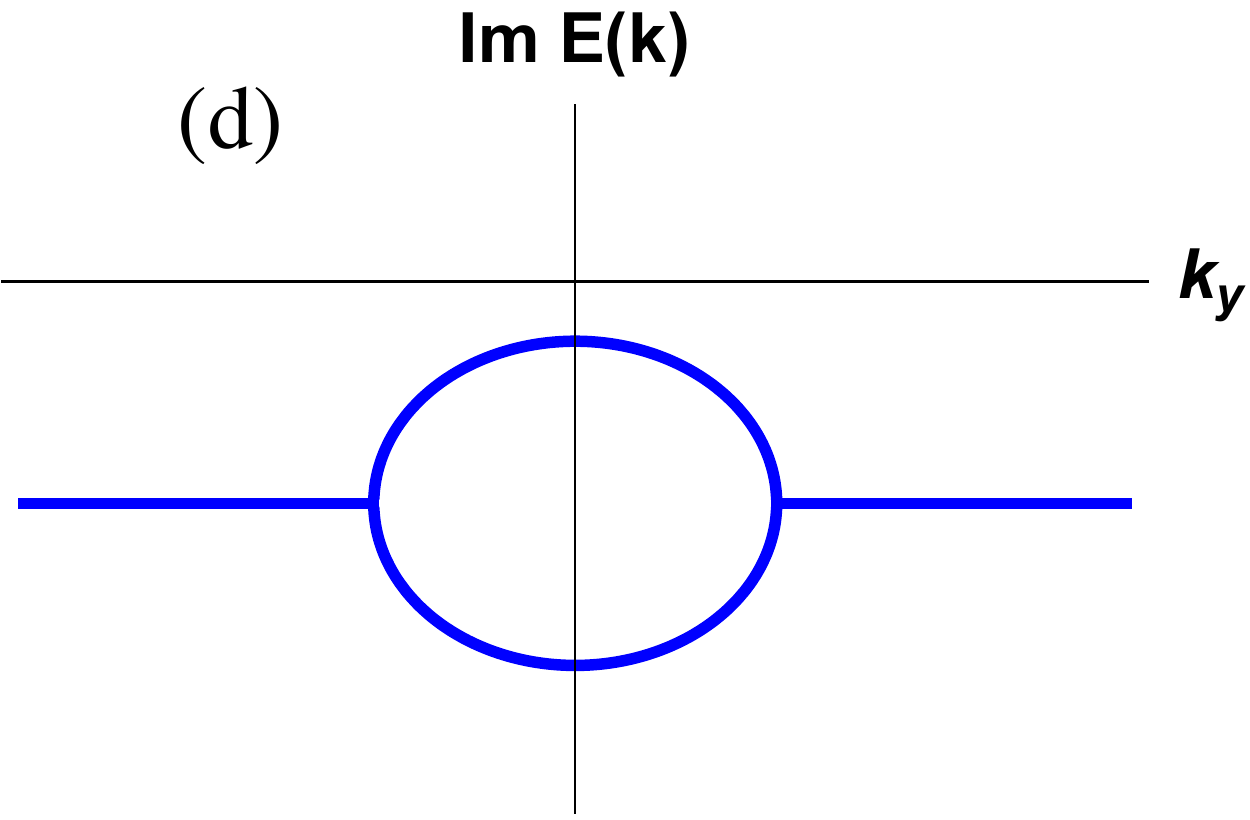}
\end{tabular}
\caption{Panel (a) shows the gapped real part of the spectrum $E(\mathbf{k})$ given by Eq. (\ref{spectrum_bulk}) at $k_x=0$ as a function of momentum $k_y$ at $\Gamma_2 /|\kappa|< \Delta$. Panel (b) shows the spectrum at the phase transition point $\Gamma_2/|\kappa| =\Delta$, at which the superconducting gap closes and then transforms into two exceptional points. Panels (c) and (d) show respectively the real and imaginary parts of the spectrum at $\Gamma_2/|\kappa|>\Delta$.
} \label{fig1}
\end{figure}

\subsection{Superconducting case}
We now consider the superconducting case where the superconducting gap is induced by proximity and analyze how
 the spectrum of 2D quasiparticles might be affected by  disorder scattering. 

{\em Weak tilt. --} In the superconducting case the spectrum is gapped at $\kappa=0$. 
At a weak tilt $0<|\kappa|<1$,  %, the superconducting gap $\Delta$ and the frequency $E$ acquire different disorder renormalizations. }
in the limit $\Gamma_1 \ll \Delta $ where the first-order Born approximation should apply, we obtain (details are presented in Appendix \ref{sec:self-energy_A2})
\begin{eqnarray}\nonumber\label{SelfEnergy_Gap}
\Sigma(E) &=& -\frac{\gamma}{2\pi v^2\sqrt{1-\kappa^2}} \ln \frac{2\Lambda}{\sqrt{\Delta^2-\frac{E^2}{1-\kappa^2}}}
 \\
&\times&\bigg[ \frac{E}{1- \kappa^2}  \left( 1 +\kappa \sigma_y\tau_z \right) -  \Delta \tau_x \bigg].
\end{eqnarray}
We can notice in this expression that the increase of disorder $\gamma$ decreases the proximity induced superconducting gap and increases the anisotropy of the dispersion. 

In the strong disorder limit, $\Gamma_1 \gg \Delta/|\kappa|^3$, we can self-consistently obtain the Fermi arc in the spectrum bounded by the exceptional points due to the imaginary part of the self-energy 
\begin{equation} 
\Sigma(0) =\frac{1-\kappa^2}{2\kappa^2}\Delta \tau_x - i \left(1+ \kappa \sigma_y \tau_z\right)\Gamma_1,
 \end{equation}
 which renders the disorder averaged Green function of quasiparticles non-Hermitian (the derivation of this result is detailed in Appendix \ref{sec:self-energy_A2}). 
This means, in particular, that disorder drives the system through a Lifshitz transition between a gapped state and a state where the superconducting gap closes at the bulk Fermi arc. 

{\em Strong tilt. --} Such a phase transition can also take place in the  strong tilt case, as we show below. Indeed, in the limit $|\kappa| > 1$ and considering that $|\kappa| k_0 \gg |E|$ and $|\kappa| k_0 \gg \Delta$, one recovers Eq. (\ref{Sigma_2}),
%\begin{eqnarray}\label{Eq10}
%\Sigma(E) = [1-Z]( E - \Delta\tau_x ) -i Z \Gamma_2 \left(1+ \frac{\sigma_y}{\kappa}\tau_z \right).~~~
%\end{eqnarray}
\begin{eqnarray}\nonumber\label{Eq10}
\Sigma(E) = -i \Gamma_2 \left(1+ \frac{\sigma_y}{\kappa}\tau_z \right) + O(E/|\kappa| k_0; \Delta/|\kappa| k_0).~~~
\end{eqnarray}
The self-energy contains a frequency independent imaginary contribution that results in an
unusual complex quasiparticles spectrum:
\begin{eqnarray}\label{spectrum_bulk}
E_{\pm}(\mathbf{k}) = \kappa vk_x - i \Gamma_2 \pm \sqrt{\Delta^2 + v^2k_y^2+\left(vk_x+i\frac{\Gamma_2}{\kappa}\right)^2 }.~~~~~
\end{eqnarray} 
We find that at $\Gamma_2 <|\kappa|\Delta$ the spectrum of the quasiparticles is gapped 
[see Fig. \ref{fig1}(a)], while in the opposite limit $\Gamma_2>|\kappa|\Delta $ the 
spectrum acquires a bulk Fermi arc bounded by two exceptional points at $\mathbf{k}_{\pm} = (0,\pm v^{-1}\sqrt{\Gamma_2^2/\kappa^2 - \Delta^2})$ as represented in  Fig. \ref{fig1}. This constitutes an example of a disorder-induced topological transition in the quasiparticles spectrum in a superconducting Dirac Hamiltonian.
The real and imaginary parts of the spectrum have square root singularity at the exceptional points. This result is analogous to the spectral properties at the Fermi arc in 2D Dirac semimetals \cite{Kozii}. 

\begin{figure}[t]
\includegraphics[width=8cm]{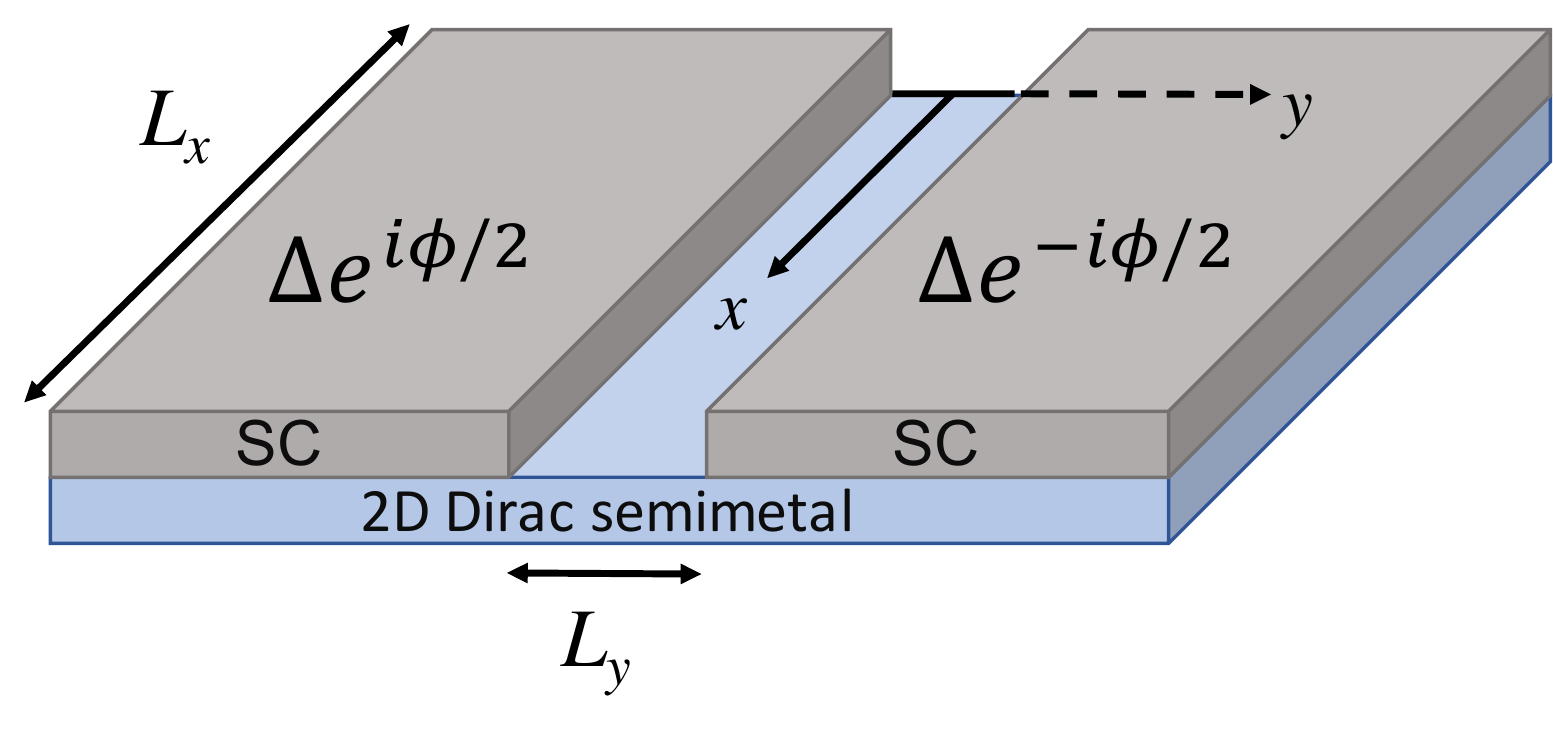}
\begin{tabular}{cc}
\includegraphics[width=4cm]{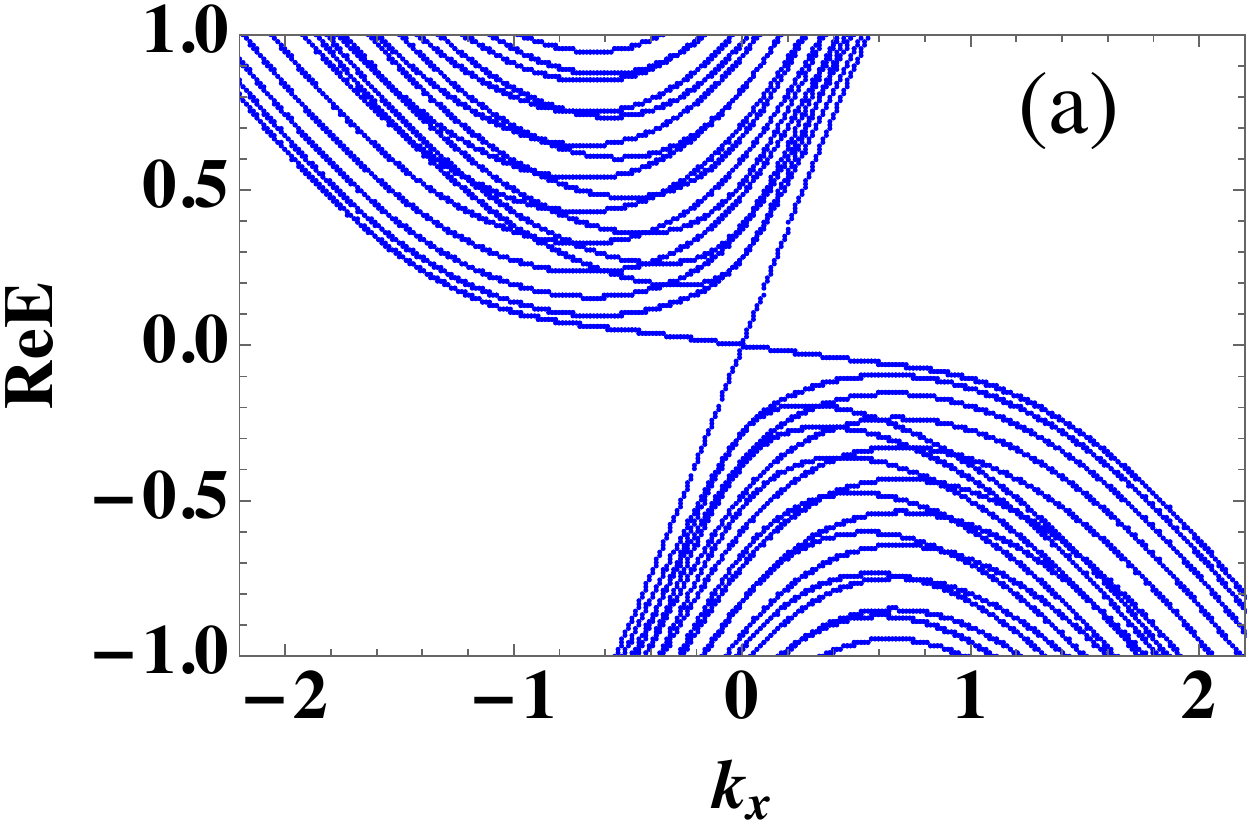} \includegraphics[width=4cm]{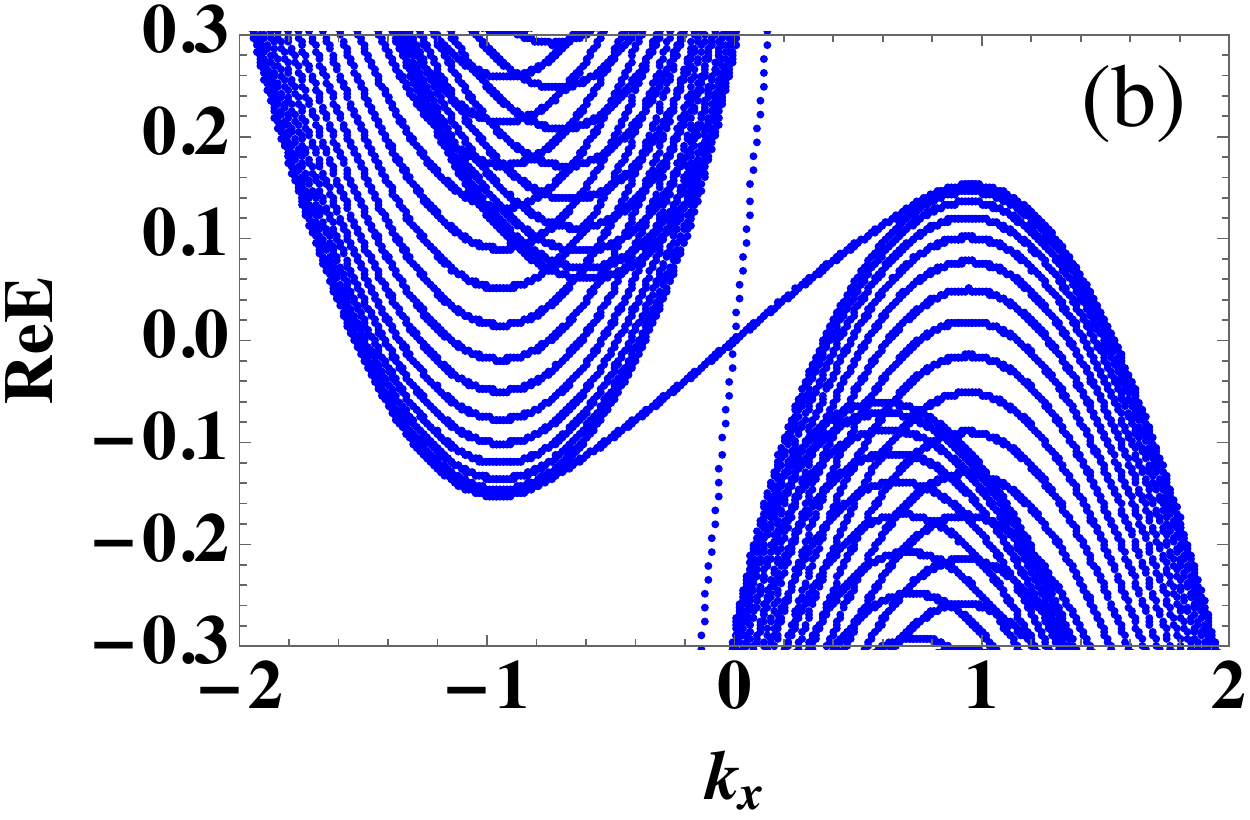}
\\  
\includegraphics[width=4cm]{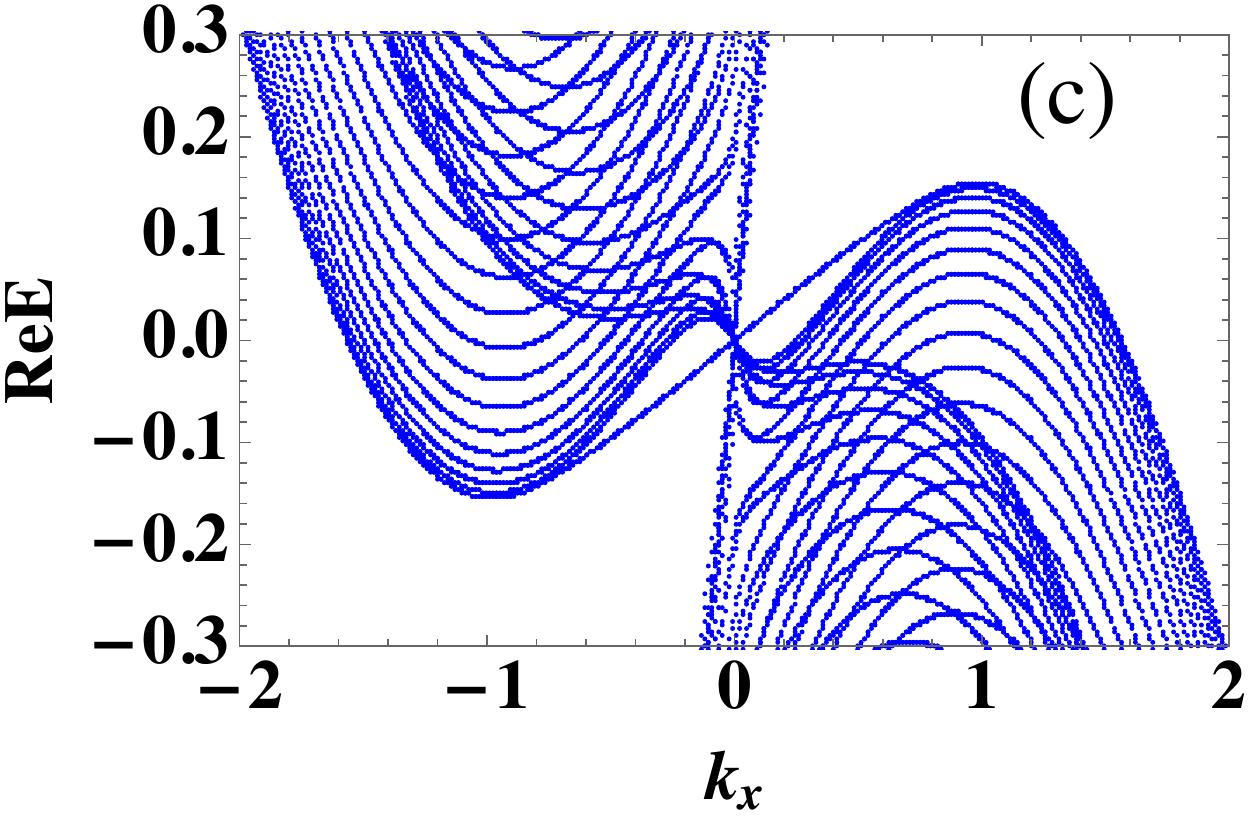}  \includegraphics[width=4cm]{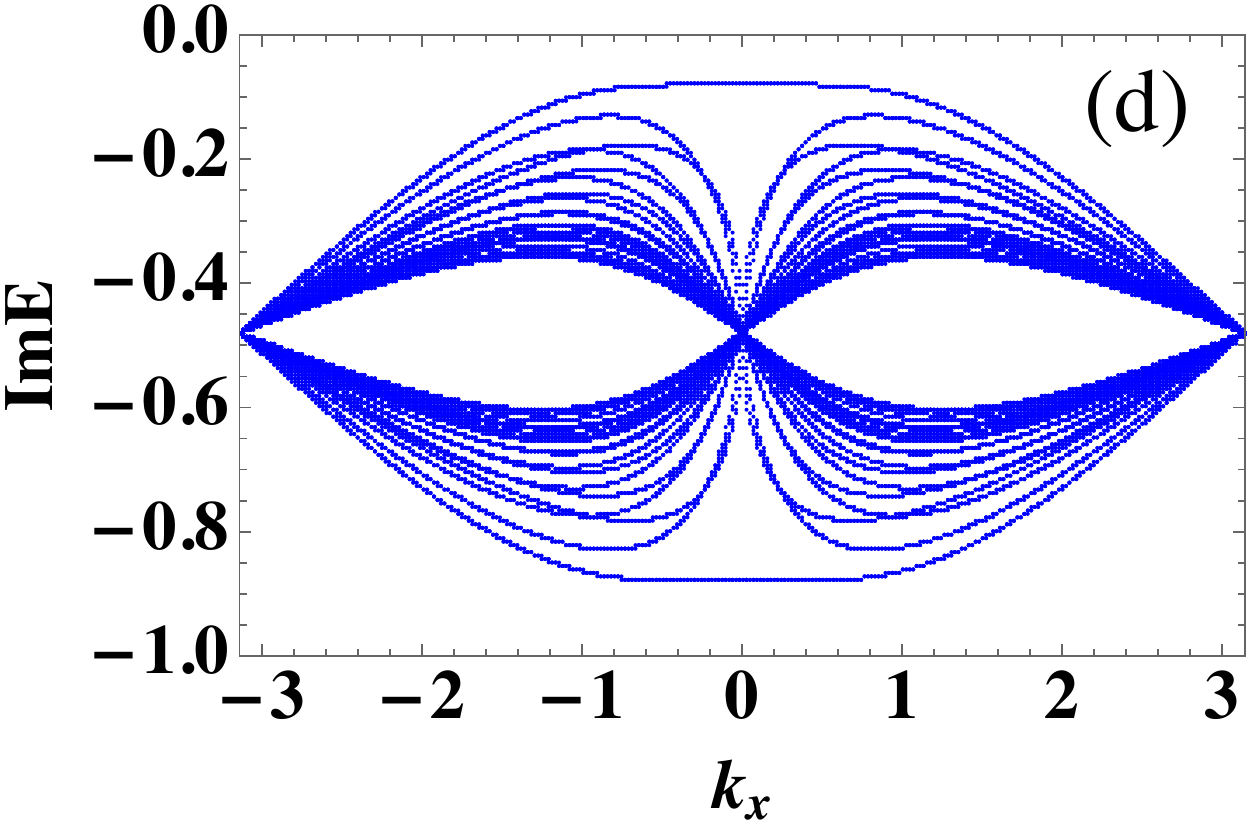}
 \\
\end{tabular}
\caption{
Top: Scheme of the Josephson junction through 2D states of the Dirac semimetal. We assume the  condition 
$L_y \ll \frac{v|(|\kappa| -1) |}{\Delta} \ll  L_x$ is met.
Bottom: Spectrum for the two-dimensional Dirac semimetal with proximity induced superconducting gap from a numerical
diagonalization of the discretized "non-Hermitian" Hamiltonian, $H+\Sigma = \kappa \sin k_x+[\sigma_x\sin k_y-\sigma_y(\sin k_x + i \gamma )]\tau_z  + (2 - \cos k_x - \cos k_y)\sigma_z - i \Gamma +\Delta\tau_x $ 
as a function of $k_x$ for a sample of finite width in the $y$-direction (here the energy is measured in the units of nearest-neighbor hopping energy, and the distance in units of the lattice constant). 
(a,b,c) Two one-way propagating chiral modes and the bulk modes, where (a) describes weak tilt, while (b) and (c) describe strong tilt cases. The increase of the scattering strength merges surface modes with the bulk. (d) The imaginary part of the dispersion. The top and bottom flat regions correspond to imaginary part of the two surface state's dispersion, which directly connect with the bulk modes. Parameters are chosen as follows: $\Delta=0.3$, (a) $\kappa =0.9$, $\Gamma=0.2$, and $\gamma = 0.18$; (b) $\kappa =1.2$, $\Gamma=0.12$, and $\gamma = 0.1$; (c), (d) $\kappa =1.2$, $\Gamma=4.8$, and $\gamma = 0.4$.} \label{fig2}
\end{figure}

{\em Surface Andreev modes. --} It is also instructive to examine Andreev modes localized at the line interface between two regions of 2D Dirac semimetal, which are proximitized to two uniform superconductors with phase difference $\phi$, as schematically shown in Fig. \ref{fig2}.
Let us focus on the limit of strong tilt $|\kappa|> 1$ where the phase transition between the gapped and the Fermi arc states can be obtained from Eq. (\ref{spectrum_bulk}). 
Following Refs. \cite{Fu_Kane}, we adopt a step function model, assuming the width of the Josephson junction to be much smaller than $v(|\kappa|-1)/\Delta$; namely we set $L_y \rightarrow 0$. 

Consider the interface along the line $y=0$, at which the proximity induced superconducting gap is given by
$
\Delta(y) = \Delta e^{-i \phi/2}
$
for $y>0$ and $\Delta(y) = \Delta e^{i \phi/2}$ for $y<0$, where gap $\Delta$ is the same in both proximitized regions. 
Note that two bulk superconductors are coupled through the 2D Dirac semimetal and there is no direct coupling between them.
At $\phi \neq 0$ there are two chiral modes, which propagate with the momentum $q$ along $x$ in the same direction, with the wave-function 
$
\Psi (q,y) \propto \exp\left(i qx - |y\sin \frac{\phi}{2} | \Delta/v\right) 
$ localized at the interface \cite{Fu_Kane}. The spectrum of these bound modes is similar to the spectrum of bulk modes given by Eq. (\ref{spectrum_bulk}), namely
\begin{equation}
E_{\pm}(q) = \kappa v q - i \Gamma_2
\pm \sqrt{\Delta^2\cos^2\frac{\phi}{2} +\left(vq+ i\frac{\Gamma_2}{\kappa}\right)^2 }.
\end{equation} 
In the special case, at $\phi = \pi$, the two chiral modes are gapless and propagate in the same direction with dispersions $E_{\pm}(q) = (\kappa\pm 1) vq - i \Gamma_2 (1\mp 1/\kappa)$.
The  edge states have a momentum independent imaginary part. Due to the strong tilt, the edge states coexist with the pockets of bulk states, but can however be
separated in energy at $|\kappa|<1$. The evolution of the spectrum as a function of the tilt is shown in Fig. \ref{fig2}. We note that there are no localized modes at the interface $x=0$ between two superconductors for $|\kappa|>1$. This result is consistent with Refs. \cite{Goerbig, Zyuzin_second}.

\begin{figure}[t]
\begin{tabular}{cc}
\includegraphics[width=4cm]{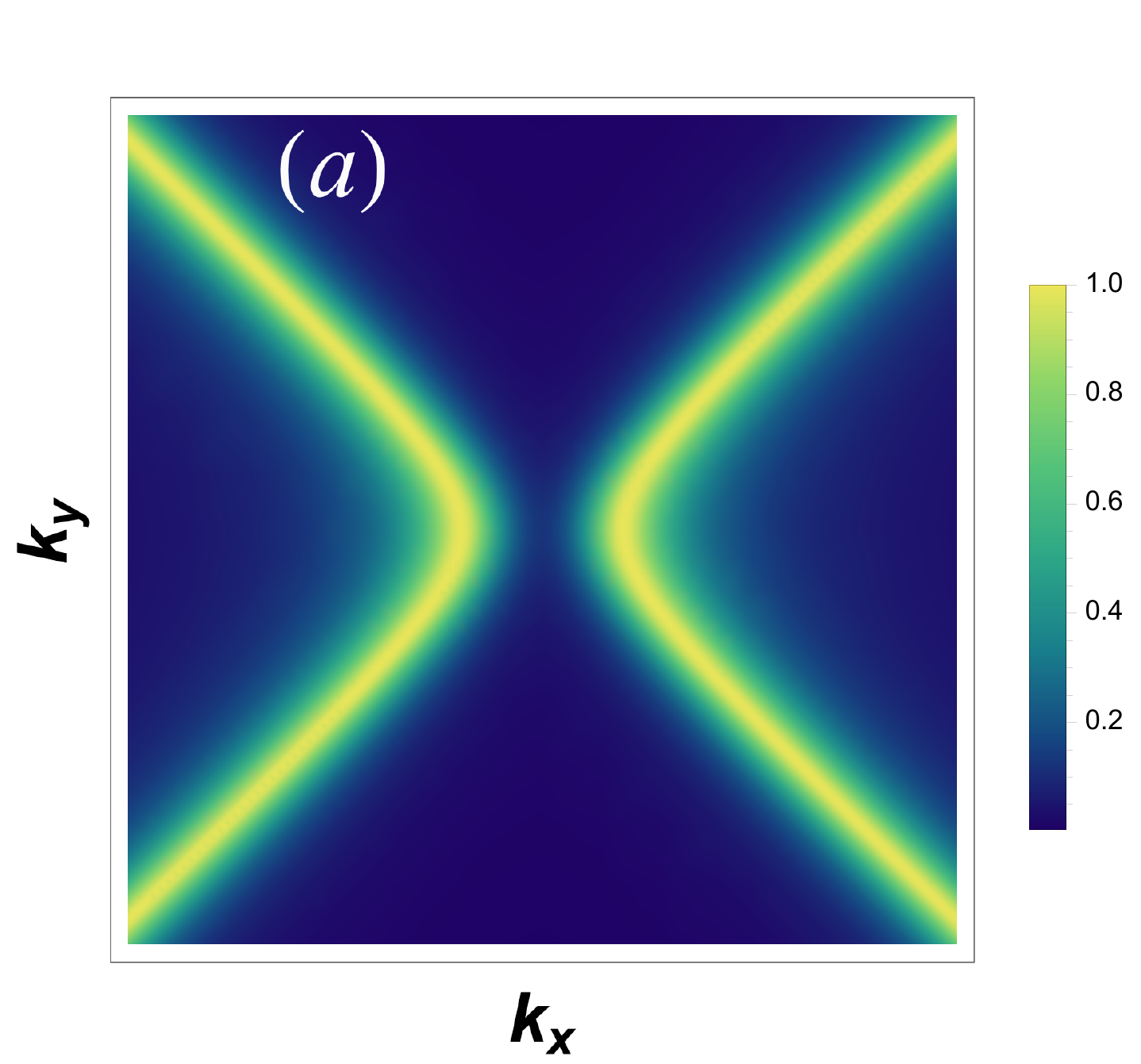} \includegraphics[width=4cm]{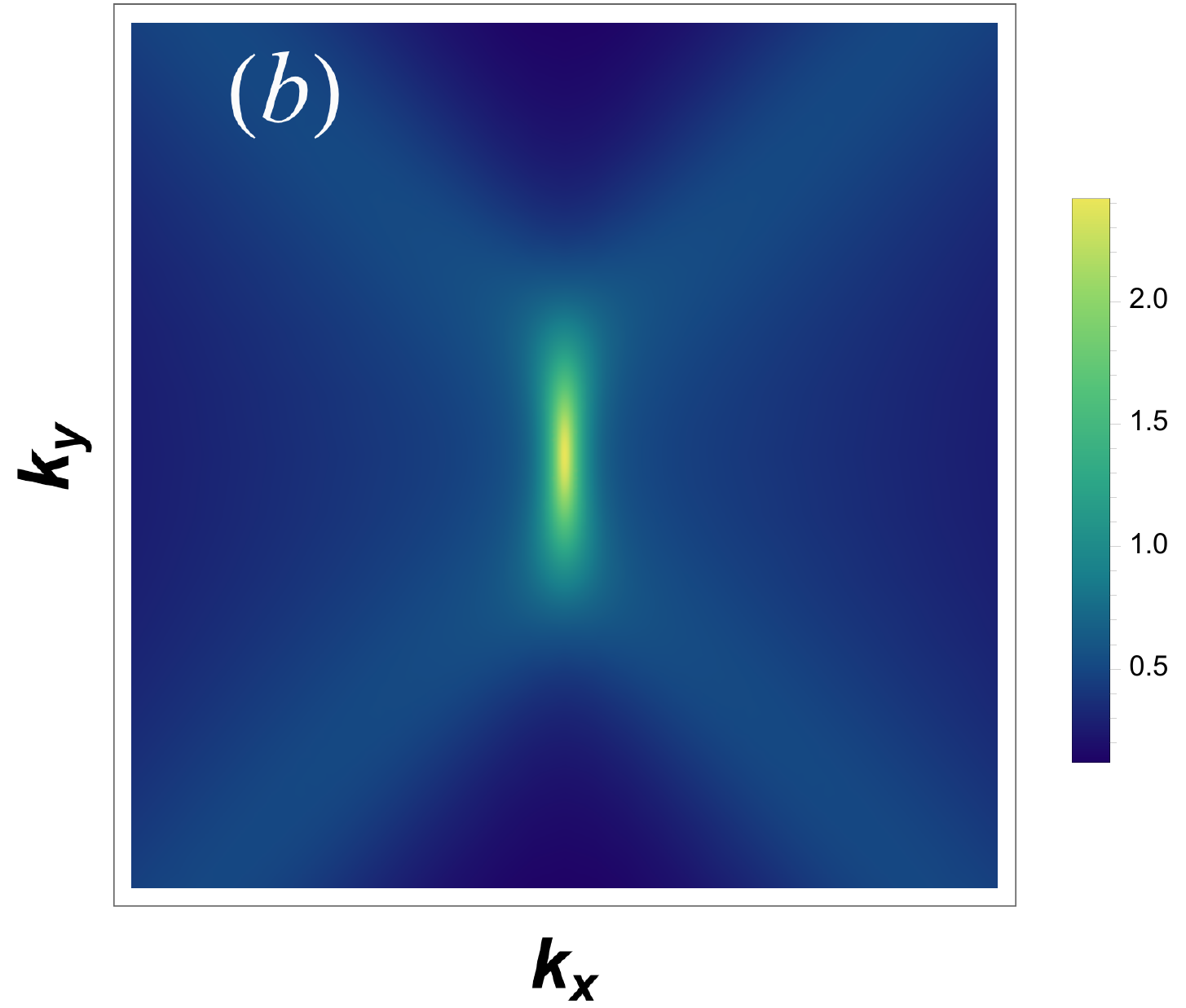}\\
\includegraphics[width=4cm]{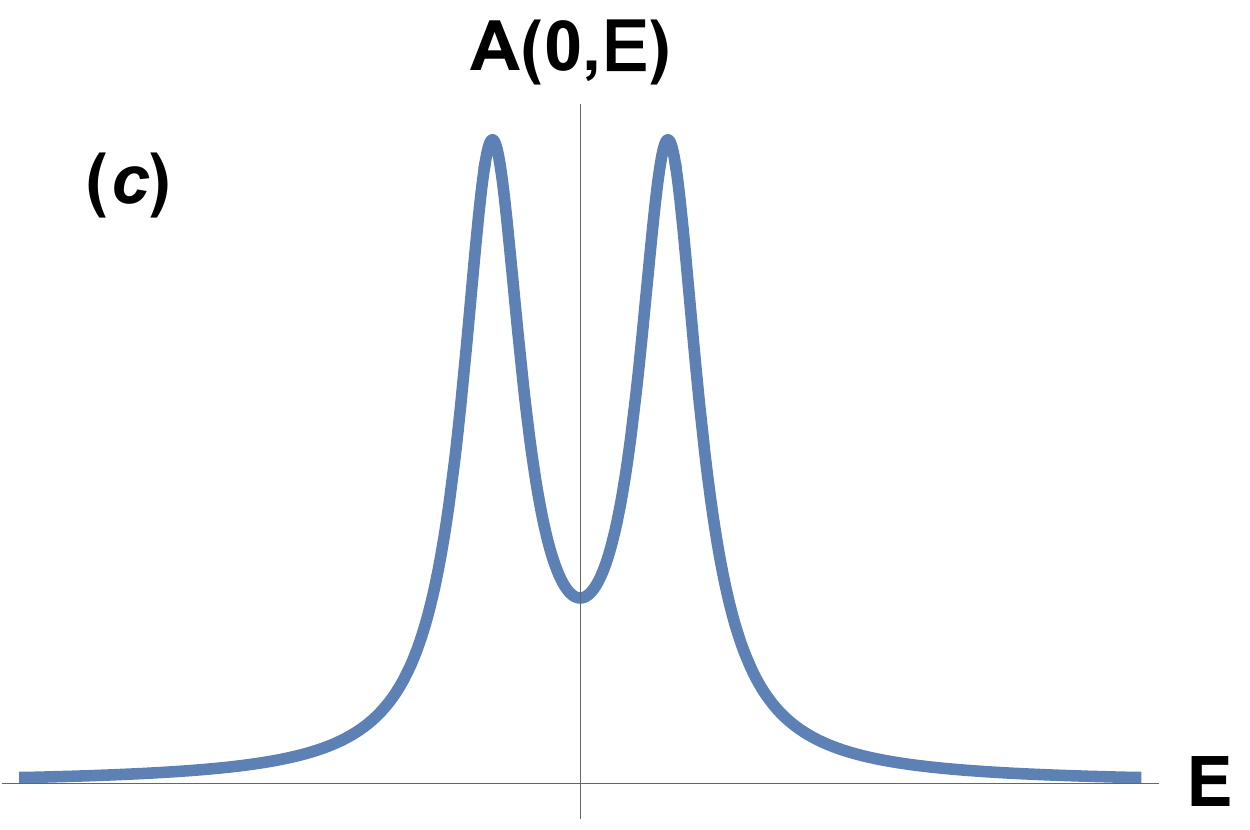} \includegraphics[width=4cm]{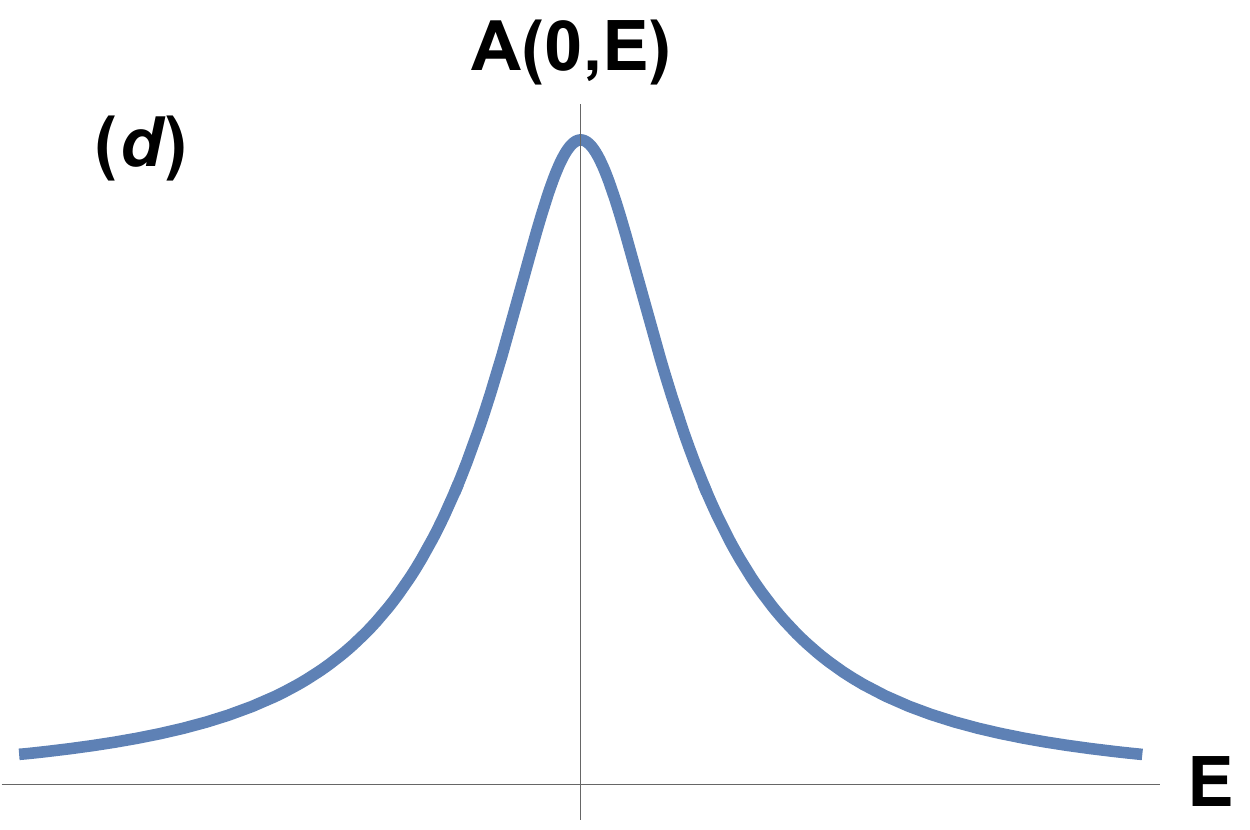}
\end{tabular}
\caption{Top: momentum resolved spectral function $A(\mathbf{k}, 0)$ at zero energy for the cases of (a) strong proximity effect, in which $\Delta/ \Gamma_2\sqrt{1+\kappa^{-2}} \approx 4>1$, and (b) weak proximity effect ($\Delta / \Gamma_2\sqrt{1+\kappa^{-2}}\approx 0.4<1$) for fixed $|\kappa| =1.3$.  The thick light green lines indicate a large spectral weight.
One sees the emergence of the bulk Fermi arc as the value of the proximity induced gap $\Delta$ decreases.
Bottom: spectral function $A(0, E)$ at zero momentum for the cases of strong (c) and weak (d) proximity effect, which can be distinguished by the number of peaks in the spectral function.}
\label{fig2new} 
\end{figure}

{\em Experimental signatures. --} Besides the specific dependence of the Andreev modes predicted for the 2D  Josephson junction depicted in Fig. \ref{fig2}, let us address the question  of how the proposed phase transition at which exceptional points emerge can be directly probed experimentally. Angle-resolved photoemission spectroscopy seems the most natural technique to probe the spectral function at the surface of a compound.
In a system with a given disorder realization, for a light beam of a width typically larger than characteristic length scale of the material (here the mean free path), the measured spectral function will be disorder averaged because different parts of the surface are probed. 
Therefore, the spectral function probed experimentally, $A(\mathbf{k},E) = -\frac{1}{\pi} \mathrm{Im} \mathrm{Tr} [E -H(\mathbf{k})-\Sigma(E)]^{-1}$ is defined as the disorder averaged retarded Green function of
quasiparticles. We plot  the spectral function at zero frequency in Figs. \ref{fig2new}(a) and \ref{fig2new}(b), where one directly observes the transition to a state hosting bulk Fermi arc as the proximity induced gap is varied. It is also instructive to comment on the frequency dependence of $A(\mathbf{k},E) $ in the limit of zero wave-vector which is related to the average density of states which can be probed by scanning tunneling microscopy (STM).  As shown in Figs. \ref{fig2new}(c) and  \ref{fig2new}(d) the spectral function has a single peak in the case of the state with the bulk Fermi arc. At strong proximity effect, $\Delta \gg \Gamma_2$, the height of two peaks located at $E_{\pm} = \pm [\Delta^2-\Gamma_2^2(1+\kappa^{-2})]^{1/2}$ is given by $A(0, E_{\pm})= 2/\pi \Gamma_2$. In the limit of weak proximity effect, $\Delta \ll \Gamma_2$, the height of the peak at zero frequency is instead given by $A(0,0)=\frac{4\Gamma_2}{\pi}[\Delta^2 + \Gamma_2^2(1-\kappa^{-2})]^{-1}$.

\section{3D nodal superconductors under weak disorder} 
Let us now consider several representative realizations of  3D topological superconductors with nodal lines or nodal points in their quasiparticles  spectrum. Notice that our discussion  equally applies to the superfluid $^3$He \cite{Volovik_book, Silaev_Volovik}. 

\subsection{Nodal-line superconductors}
The BCS Hamiltonian of a characteristic nodal line superconductor (our results can be mapped to a polar phase of superfluid $^3$He) in the presence of the supercurrent flow with velocity $\boldsymbol{v}_s$ is given by \cite{Volovik_book}
\begin{equation}
H(\mathbf{k}) = \frac{k^2-k_F^2}{2m} \tau_z + \boldsymbol{v}_s\cdot \mathbf{k} + \Delta n_z \tau_x 
\end{equation}
where $\mathbf{n} = \mathbf{k}/k$ is the unit vector in the direction of momentum, $k_F$ is the Fermi momentum, $m$ is the effective mass, and $\Delta(\boldsymbol{v}_s)$ is the gap in the spectrum 
(the gap is suppressed with the increase of supercurrent). 
The spectrum of quasiparticles at $|\boldsymbol{v}_s|=0$ has a nodal ring defined by the equation $k_x^2+k_y^2 = k_F^2$ in the plane $k_z=0$. We first consider a case of the supercurrent flowing along the $z$-axis. In this situation the Fermi surface appears above the Lifshitz transition occurring at $|\boldsymbol{v}_s|k_F = \Delta$. 

The self-energy due to scattering on the scalar disorder can be found self-consistently. 
Consider first the situation where the tilt is along the $z$-axis, namely we set for concreteness $\boldsymbol{v}_s = v_s(0,0,1)$.
The retarded self-energy taken at zero frequency within the first order in powers of $|v_s|k_F/ \Delta < 1$ is given by
\begin{equation}\label{polar_z}
\Sigma(0) =  - i\Gamma \left[ 1 + \frac{v_sk_F}{\Delta}\tau_x \right],
\end{equation}
where  $\Gamma = \Delta \mathrm{csch}(2\tau \Delta)$ and $1/2\tau = \pi \nu \gamma $ is the scattering rate (details of this result are shown in Appendix \ref{sec:self-energy_2}). The first term in Eq. (\ref{polar_z}) is well known \cite{Gorkov, Mineev_book} and implies that an infinitesimally weak scattering smears the nodal ring in three dimensions.

\begin{figure}[t]
\begin{tabular}{cc}
\includegraphics[width=4cm]{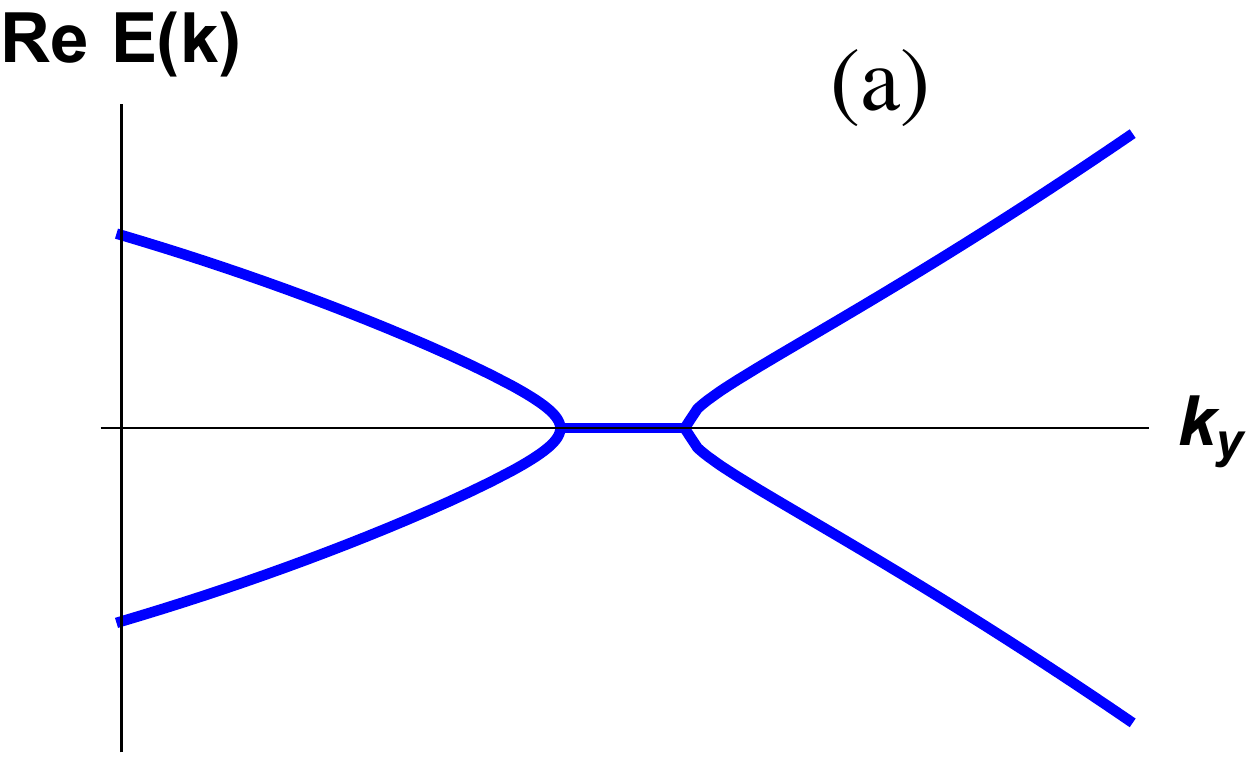} \includegraphics[width=4cm]{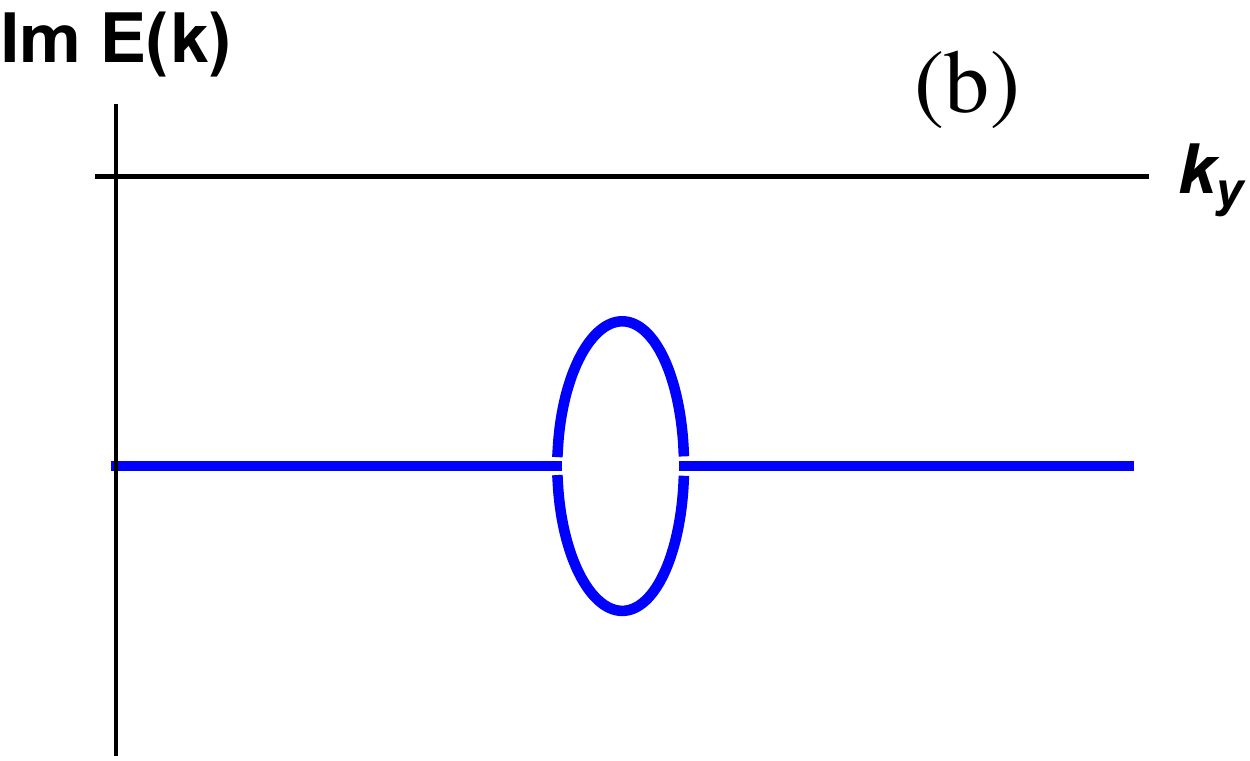}
\end{tabular}
\caption{
Panels (a) and (b) show respectively the real and imaginary parts of the spectrum $E(\mathbf{k})$, given by Eq. (\ref{spectrum_he_1}), as a function of momentum $k_y$  at $k_x=k_z=0$ for the disordered nodal-ring superconductor (or polar phase of the superfluid $^3$He) with the supercurrent applied along the $z$-axis. 
The interplay of the supercurrent and disorder splits the nodal-ring into a region where the real part of the spectrum is zero, bounded with the outer and inner exceptional circumferences. 
The spectrum has a square root singularity at the exceptional rings.} \label{fig3}
\end{figure}

We show that this smearing becomes anisotropic due to the tilt, which gives rise to the off-diagonal terms in the self-energy matrix in
 Eq. (\ref{polar_z}), with the sign being defined by $v_s$, i.e., by the direction of the supercurrent. 
Indeed, taking into account the self-energy in Eq. (\ref{polar_z}), we can write the spectrum of quasiparticles in the plane $k_z=0$ in the following form:
\begin{eqnarray}\label{spectrum_he_1}
E_{\pm}(\mathbf{k}) = - i \Gamma \pm \sqrt{\left(\frac{k_x^2+k_y^2-k_F^2}{2m} \right)^2 - \left(\frac{v_sk_F}{\Delta}\Gamma \right)^2}.~~~~
\end{eqnarray} 
The nodal ring in the disordered superconductor extends into a ring of finite width (called a Fermi ribbon) with the inner and outer radiuses defined by 
$
k_{\pm} = k_F\sqrt{1 \pm \frac{v_sk_F}{\Delta}\frac{\Gamma}{\mu}},
$
where
the matrix of the disorder averaged Green function is defective. This result is analogous to the theoretically predicted exceptional rings in the nodal-line semimetals \cite{Zyuzin_second, Johan_Emil_forth, Yang_fith, Wang_sixth,  Moors, Okugawa, Budich}. The spectral properties of the Fermi ribbon is similar to that of the Fermi arc discussed in the previous section. The dispersion is shown in Fig. \ref{fig3}. 
The nodal-loop superconductor has so-called "drum-head" states localized at any surfaces not laying parallel to the $z$-axis, \cite{Burkov_Hook}.
Due to the smearing of the nodal-line the spectral region of these edge states vanishes.

Let us  also contrast these results with the situation in which  the supercurrent is applied within the $k_z=0$ plane.
The supercurrent tilts the nodal ring and gives rise to  Fermi surface pockets connected by 
two pseudo-Weyl points, which are located at $\pm k_F \boldsymbol{v}_s\times \hat{\mathbf{z}}/|\boldsymbol{v}_s|$, \cite{Volovik_uspekhi}. Here the density of states is finite even at infinitesimally small $|v_s|$ and is given by $\nu |\boldsymbol{v}_s|k_F/\Delta$, in which $\nu = mk_F/2\pi^2$ is the density of states at the Fermi level for one spin projection in the normal state of the system.
At $|v_s|k_F/\Delta \gg \tau \Gamma$, the self-energy at zero frequency is linear $\Sigma = -i |v_s|k_F/2\tau \Delta$ in $v_s$, while in the limit $|v_s| k_F/\Delta \rightarrow 0$  it reaches the value $\Sigma=-i\Gamma[1 + \mathcal{O}(|v_s|^2 k_F^2/\Delta^2) ]$. We did not find any exceptional points in this case because $\mathbf{v}_s\cdot n_z =0$. 

\subsection{Point-node superconductors}
Let us finally comment on the effect of the interplay between the supercurrent and disorder scattering in a 3D Weyl superconductor with  nodal points in their quasiparticles spectrum.
The BCS Hamiltonian describing a Weyl superconductor (or equivalently the A-phase of  $^3$He) is given by
\begin{equation}
H(\mathbf{k}) = \frac{k^2-k_F^2}{2m} \tau_z + \mathbf{v}_s\cdot \mathbf{k} + \Delta (n_y \tau_x - n_x \tau_y ).
\end{equation}
At $|\mathbf{v}_s|=0$, the spectrum of quasiparticles has two Weyl nodes at $\mathbf{k} =(0,0,\pm k_F)$. The surface of the superconductor might host Andreev-Majorana localized chiral modes. The dispersion of surface modes has the form of the Fermi arc, which connects the projections of the bulk Weyl points to the surface; for a review see Ref. \cite{Silaev_Volovik}. 

Consider qualitatively the situation where the supercurrent flows parallel to the $x$-axis, $\mathbf{v}_s=v_s(1,0,0)$. At $|v_s|k_F<\Delta$, similarly to 3D Dirac-Weyl semimetals, weak disorder satisfying $\tau\Delta <\pi/4$ does not lead to a finite imaginary part of the self energy at zero frequency and hence does not form any exceptional lines. 

\section{Discussion and Conclusions} 
In this paper we have studied the effect of weak scalar disorder on the band structure of nodal superconductors. We have argued that the nodes in the anisotropic superconducting gap in the presence of weak disorder 
may be replaced by  Fermi arcs or 2D Fermi  areas bounded by  exceptional points or exceptional lines, respectively. At these exceptional points or lines  the quasiparticles Green function   becomes defective.
Here we have analyzed the smearing of the nodes in the superconducting gap in the presence of  scalar disorder within the self-consistent Born approximation.
Going beyond this approximation and  taking into account a more general form for the disorder \cite{Aleiner_Efetov} shall be addressed in the  future.

We believe that the proposed non-Hermitian superconducting phase might be probed in several materials, such as the quasi-2D organic conductor 
$\alpha \mathrm{- (BEDT-TTF)_2I_3}$ salt \cite{Organic_Conductor, Goerbig_2d_tilt} and (001) surface states of the crystalline insulator $\mathrm{SnTe}$
\cite{Tanaka_SnTe}, which host 2D massless Dirac fermions with anisotropic dispersion. The proximity induced superconducting gap in these structures might be established experimentally. 
Nodal phases with the zeros in the energy spectrum are known to exist in superfluid $^3$He-A (for a review see \cite{Silaev_Volovik}). Signatures of nontrivial topological
nodal superconductivity were observed in Cu$_x$Bi$_2$Se$_3$, in non-centrosymmetric heavy fermion systems, and in cuprate-based superconductors \cite{Sato_Ando_review}.

Finally, it would be interesting to extend our work on the interplay between  proximity induced superconductivity and disorder in systems with  triple Dirac points \cite{Bradlyn}.  Therein,  higher-order non-Hermitian degeneracies with cubic-root singularities at the nontrivial exceptional curves are expected  \cite{Third_order_Dirac_EP}.

{\bf Acknowledgements}
We thank G. Volovik for important comments. A.A.Z. acknowledges the hospitality of the Universit\'e Paris-Sud and Pirinem School of Theoretical Physics as well as the support by the Academy of Finland.

\appendix
\section{Calculation of the self-energy in two dimensions.}
\subsection{Normal case $\Delta=0$}\label{sec:self-energy_A1}
Let us calculate the disorder induced self-energy correction for the case of a two-dimensional semimetal. The Hamiltonian describing such a system 
reads
\begin{equation}
H(\mathbf{k}) = \kappa v k_x + v(k_y\sigma_x -k_x\sigma_y).
\end{equation}
The self-consistent equation for the retarded self-energy within the first Born approximation is given by
\begin{equation}
\Sigma(E) = \gamma\int \frac{d^2k}{(2\pi)^2}[E - H(\mathbf{k}) - \Sigma(E) ]^{-1}.
\end{equation}
We search for the solution of this equation in the form
\begin{equation}
\Sigma = t + d \sigma_y + \epsilon \sigma_x, 
\end{equation}
where $t=t_1- i t_2 $, $d = d_1- id_2 $, and $\epsilon$ are complex functions of the energy $E$ such that $t_{1,2}$ and $d_{1,2}$ are  real and satisfy $t_2>|d_2|\geq 0$ because we are dealing with a retarded self-energy.
Since the tilt is along the $k_x$ direction, the $\epsilon \sigma_x$ contribution in the self-energy vanishes to guarantee its momentum independence. 
This point can also be checked explicitly.
Hence
\begin{equation}\label{Sigma_wt}
\Sigma = \gamma \int \frac{d^2k}{(2\pi)^2} \frac{ E-t -\kappa v k_x +\sigma_x(vk_y+\epsilon)-\sigma_y(vk_x-d)}{( t -E+ \kappa v k_x)^2 - (vk_y +\epsilon)^2 - (vk_x - d)^2}.
\end{equation}
From hereon we consider $E=0$ to make progress with the algebra. The poles in \ref{Sigma_wt} can be found from the equation
\begin{equation}
\left( vk_x - \frac{d+ \kappa t}{1-\kappa^2}\right)^2 = -\frac{(vk_y+\epsilon)^2}{1-\kappa^2} + \left( \frac{t+\kappa d}{1-\kappa^2}\right)^2.
\end{equation}
When $|\kappa|<1$, we note that provided
\begin{equation}
|d_2+ \kappa t_2| < \left|\mathrm{Re}\sqrt{(vk_y+\epsilon)^2(1-\kappa^2) - \left( t+\kappa d\right)^2}\right|
\end{equation}
the integration over momentum $k_x$ yields 
\begin{eqnarray}\label{Intermediate}\nonumber
\Sigma &=& \frac{\gamma}{1-\kappa^2} \bigg[ t -d \sigma_y+ (\kappa+\sigma_y) \frac{d+\kappa t}{1-\kappa^2} - \sigma_x(w+\epsilon)\bigg]\\
&\times& \int_{-\Lambda}^{\Lambda} \frac{dw}{4\pi v^2}\left[ \frac{(w+\epsilon)^2}{1-\kappa^2} - \left(\frac{t +\kappa d }{1-\kappa^2} \right)^2 \right]^{-1/2},
\end{eqnarray}
where $\Lambda$ is the energy cut-off and $w \equiv vk_y$. It can be now seen that the $\sigma_x$ term is small $\propto O(\epsilon/\Lambda)$ and will be neglected in what follows. 

The integration over $w$ in Eq. (\ref{Intermediate}) after some simplifications results in two equations for $t$ and $d$:
\begin{eqnarray}\label{td}\nonumber
t&=&\frac{\gamma }{2\pi v^2} \frac{t+ \kappa d}{(1-\kappa^2)^{3/2}} \ln \frac{2\Lambda\sqrt{1-\kappa^2}}{i(t +\kappa d )},\\
d&=& \frac{\gamma \kappa }{2\pi v^2} \frac{t+ \kappa d}{(1-\kappa^2)^{3/2}} \ln \frac{2\Lambda\sqrt{1-\kappa^2}}{i(t +\kappa d )}.
\end{eqnarray}
Noting that $d = \kappa t $, we obtain for $t \neq 0$
$$
1 = \frac{\gamma }{2\pi v^2} \frac{1+ \kappa^2}{(1-\kappa^2)^{3/2}} \ln \frac{2\Lambda\sqrt{1-\kappa^2}}{it(1 +\kappa^2 )},
$$
which gives $\Sigma = -i (1+\kappa\sigma_y) \Gamma_1$, where 
\begin{equation}
\Gamma_1  = 2\Lambda \frac{\sqrt{1-\kappa^2}}{1+\kappa^2} \exp\bigg[-\frac{2\pi v^2}{\gamma}\frac{(1-\kappa^2)^{3/2}}{1+\kappa^2}\bigg].
\end{equation}
This is the expression given in Eq. (\ref{Gamma_1}) of the main text.

When $|\kappa|>1$ it is enough to consider the self-energy within the first Born approximation similarly as it was shown for the nodal-line semimetal in \cite{Zyuzin_second}.
Integrating first over $k_y$,
\begin{eqnarray}\nonumber
\Sigma &=& -i\frac{\gamma}{8\pi } \sum_{s=\pm}\int_{-k_0}^{k_0} dk_x \int dk_y \left[1 + s\frac{ k_x}{k}\sigma_y \right]  \\
&\times&\delta(\kappa v k_x - s vk)= -\frac{i\gamma}{2\pi v} \left(1+ \frac{\sigma_y}{\kappa} \right)\frac{|\kappa|k_0}{\sqrt{\kappa^{2}-1}},~~~~~
\end{eqnarray}
where $k_0(\kappa)$ is the momentum cutoff on the $k_x$-axis describing the width of the electron and hole pockets, which might depend on the tilt parameter $\kappa$.
This is the expression given in Eq. (\ref{Sigma_2}) of the main text.

\subsection{Superconducting case}\label{sec:self-energy_A2}
The BdG Hamiltonian describing the low-energy states in the system  in the presence of the proximity-induced superconducting gap reads
\begin{equation}
H(\mathbf{k}) = \kappa v k_x + v(k_y\sigma_x -k_x\sigma_y) \tau_z + \Delta \tau_x,
\end{equation}
where $\Delta$ can be considered to be real and positive without loss of generality. In order to obtain an analytical solution for $\Sigma$, we again proceed by considering two limiting cases of weak and strong tilts, $|\kappa| < 1$ and $|\kappa| > 1$.

When $|\kappa| < 1$ and $\Delta \gg \Gamma_1$, we can use the first-order Born approximation. For convenience we transform to Matsubara frequencies $E+i0^+ \rightarrow i\omega_n$. The self-energy $\Sigma \rightarrow \Sigma_{\omega_n}$ then reads 
\begin{align}\nonumber
&\Sigma_{\omega_n} = \gamma\int \frac{d^2k}{(2\pi)^2} \frac{i\omega_n - \kappa v k_x + v(k_y\sigma_x-k_x\sigma_y)\tau_z-\Delta\tau_x}{(i\omega_n - \kappa v k_x)^2 - v^2k^2 - \Delta^2}\\
&= \frac{i\gamma}{2(1-\kappa^2)} \int \frac{d^2k}{(2\pi)^2} \frac{1}{\sqrt{\lambda}} \sum_{s=\pm} s \frac{i\omega_n-\Delta\tau_x - v k_x(\kappa +\sigma_y\tau_z)}{vk_x +i\omega_n\frac{\kappa}{1-\kappa^2} - is\sqrt{\lambda }},
\end{align}
where 
\begin{equation}
\lambda=\frac{k_y^2+\Delta^2}{1-\kappa^2} +\frac{\omega_n^2}{(1-\kappa^2)^2}. 
\end{equation}
Provided $\mathrm{Re}\sqrt{\lambda}> \frac{|\omega_n\kappa|}{1-\kappa^2}$ (in the opposite case the integral is zero) after the integration over $k_x$ one obtains
\begin{eqnarray}
\Sigma_{\omega_n}
= -\frac{\gamma}{2\pi v^2}\int_0^{\Lambda} dw \frac{i\omega_n-\Delta\tau_x + (\kappa+\tau_z\sigma_y) \frac{i\omega_n \kappa}{1-\kappa^2}}{\sqrt{(w^2 +\Delta^2)(1-\kappa^2)+ \omega_n^2}},~~~~~~~
\end{eqnarray}
where $\Lambda$ is the energy cutoff. The integration over $w$ gives
\begin{eqnarray}\nonumber
\Sigma_{\omega_n} &=& -\frac{\gamma}{2\pi v^2\sqrt{1-\kappa^2}} \ln \frac{2\Lambda}{\sqrt{\Delta^2+\frac{\omega_n^2}{1-\kappa^2}}}
 \\
&\times&\bigg[ \frac{i\omega_n}{1-\kappa^2}  \left( 1 + \kappa \sigma_y\tau_z \right) - \Delta\tau_x \bigg].
\end{eqnarray}
Notice that there is no contribution proportional to $\sigma_x$. After the transformation $i\omega_n \rightarrow E+i0^+$ and $\Sigma_{\omega_n}\rightarrow \Sigma$, one obtains the expression in Eq. (\ref{SelfEnergy_Gap}) of the main text.

When $|\kappa| < 1$ and $|\kappa|\Gamma_1\gg \Delta $, following the same reasoning as in  Appendix \ref{sec:self-energy_A1}, 
we now search for the self-energy in the form
\begin{equation}
\Sigma = t + d \sigma_y \tau_z + S\tau_x, 
\end{equation}
where $S=S_1- iS_2$.
The integrand in the self-consistent equation for the self-energy now 
reads
\begin{eqnarray}\nonumber
\Sigma &=& \gamma\int \frac{d^2k}{(2\pi)^2} \left[E-t- \kappa v k_x - (vk_x-d)\sigma_y\tau_z - (\Delta+S)\tau_x \right]\\\nonumber
&\times&\left[(t -E+ \kappa v k_x)^2 - v^2k_y^2 - (vk_x-d)^2-(\Delta+S)^2\right]^{-1}\\
\end{eqnarray}
and can be linearized in powers of $\Delta$.
Performing the same derivations as were done in Appendix \ref{sec:self-energy_A1}, we obtain an additional equation  in  addition to Eq. (\ref{td}) in order to determine $S$ at $E=0$: 
\begin{equation}
S = \frac{\gamma (\Delta+ S)}{2\pi v^2\sqrt{1-\kappa^2}}\ln\frac{2\Lambda \sqrt{1-\kappa^2}}{i(t+\kappa d)}.
\end{equation}
Together with Eq. (\ref{td}), this gives
\begin{equation}
S= (\Delta + S) \frac{1-\kappa^2}{1+\kappa^2}.
\end{equation}
Therefore, $S = \Delta (1-\kappa^2)/2\kappa^2$ and
\begin{equation}
\Sigma = \frac{1-\kappa^2}{2\kappa^2} \Delta\tau_x -i (1+\kappa\sigma_y\tau_z) \Gamma_1.
\end{equation}

In the limit of a strong tilt, when $|\kappa|>1$, we have a  finite density of states at the Dirac point and we can use  the first-order Born approximation to calculate the self-energy,
\begin{eqnarray}\label{Sigma_E1}
\Sigma(E) = \gamma\int \frac{d^2k}{(2\pi)^2} \frac{E+ i\delta - (\kappa  +\sigma_y\tau_z) v k_x - \Delta\tau_x}{(E+ i\delta - \kappa v k_x)^2 - v^2k^2 - \Delta^2},~~~~~~~
\end{eqnarray}
where the integral over $k_x$ is taken in the region $k_x \in [-k_0, k_0]$ to qualitatively account for the width of the electron-hole pockets at $|\kappa| >1$. It is instructive to rewrite Eq. (\ref{Sigma_E1}) as 
\begin{eqnarray}\label{Intermediate}
\Sigma(E) = \gamma\int \frac{d^2k}{(2\pi)^2} \frac{E+ i\delta - (\kappa  +\sigma_y\tau_z) v k_x - \Delta\tau_x}{M-v^2k_y^2+ (E-  \kappa vk_x) i\delta},~~~~~~~
\end{eqnarray}
where
\begin{equation}
M = v^2(\kappa^2 - 1)[(k_x-p)^2 - q], 
\end{equation}
in which $p = \kappa E/(\kappa^2-1)v$ and $q = (\frac{E^2}{\kappa^2-1} + \Delta^2)/(\kappa^2 - 1)v^2$, and note that
\begin{equation}
\mathrm{sign}[E-  \kappa vk_x] = -\mathrm{sign} \left[k_x - p + \frac{E/(\kappa v)}{\kappa^2-1}\right].
\end{equation}
We can integrate in Eq. (\ref{Intermediate}) over $k_y$ separately in two regions, $M<0$ and $M>0$, taking into account that $\mathrm{sign}(E-  \kappa vk_x)$ does not depend on $k_y$. 
For $M>0$ the integration over $k_y$ gives:
\begin{align}\nonumber
&\Sigma(E) = -i\gamma\int_{-k_0}^{k_0} \frac{d k_x}{4\pi v} [E - \Delta\tau_x - (\kappa  +\sigma_y\tau_z) v k_x] \\
&\times \frac{\mathrm{sign}(E-  \kappa vk_x)}{\sqrt{M}} = - \frac{i \gamma }{2\pi v} \frac{|\kappa| k_0 }{\sqrt{\kappa^2-1}}\left( 1+ \frac{\sigma_y}{\kappa}\tau_z\right).
\end{align}
This term is purely imaginary since we neglect small corrections $\propto p/k_0 \ll 1$. At $M<0$ we obtain
\begin{align}\nonumber
&\Sigma(E) = -\gamma\int_{-\sqrt{q}}^{\sqrt{q}} \frac{d k_x}{4\pi v^2} \frac{E - \Delta\tau_x - (\kappa  +\sigma_y\tau_z) v (k_x+p)}{\sqrt{\kappa^2-1}\sqrt{q-k_x^2}} \\
&= \frac{-\gamma}{4v^2\sqrt{\kappa^2-1}}\left[ E-\Delta\tau_x - (\kappa +\sigma_y \tau_z)\frac{\kappa E}{(\kappa^2-1)}\right] .
\end{align}
We arrive at the expression for the self-energy,
\begin{eqnarray}\nonumber
\Sigma(E) &=&   -\frac{\gamma }{4v^2\sqrt{\kappa^2 - 1}} \left[ \frac{E}{1-\kappa^2}(1+\kappa \sigma_y\tau_z) - \Delta\tau_x \right]\\
&-&\frac{i\gamma}{2\pi v} \left(1+ \frac{\sigma_y}{\kappa}\tau_z \right)\frac{|\kappa|k_0}{\sqrt{\kappa^{2}-1}}.
\end{eqnarray}
Finally, we note that $|\kappa| k_0 \gg \Delta, |E|/(\kappa^2-1)$ and write
\begin{equation}\nonumber
\Sigma(E) =-\frac{i\gamma}{2\pi v} \left(1+ \frac{\sigma_y}{\kappa}\tau_z \right)\frac{|\kappa|k_0}{\sqrt{\kappa^{2}-1}},
\end{equation}
which is given in Eq. (\ref{Eq10}) of the main text.

\section{Calculation of the self-energy in three dimensions.} \label{sec:self-energy_2}
Let us consider a nodal-line superconductor in the presence of a supercurrent flow with velocity $\mathbf{v}_s = v_s(0,0,1)$ parallel to the $z$-axis. The BdG Hamiltonian is given by
\begin{equation}
H = v_s k_z + \xi \tau_z + \Delta n_z\tau_x,
\end{equation} 
where $\xi = k^2/2m -\mu$ and $n_z= k_z/k$ is the unit vector in the direction of $\mathbf{k}_z$. We focus on the limit $|v_s|k_F \ll \Delta $. The nodal-ring is defined by $\xi=0, k_z=0$. 
The equation for the self-energy is given by
\begin{equation}
\Sigma = -\gamma \int\frac{d^3k}{(2\pi)^3} \frac{v_s k_z+t -\xi\tau_z + (\Delta n_z+S)\tau_x}{(v_s k_z  + t )^2-\xi^2-(\Delta n_z +S)^2}.
\end{equation}
We neglect the contributions to $\Sigma$ with $\tau_z$ matrix, assuming $\mu \gg \mathrm{Re Tr}(\tau_z\Sigma)$. Integration over $\xi$, 
\begin{equation}
\int \frac{d^3k}{(2\pi)^3}F(\mathbf{k}) \rightarrow \nu \int_{-\infty}^{\infty} d\xi \int_{0}^{\pi} \sin\theta \frac{d\theta}{2} F(\xi, \theta),
\end{equation}
where $\nu = mp_F/2\pi^2$ is the density of states in the normal-metal state, we obtain
\begin{equation}
\Sigma = \frac{i}{2\tau} \int_{-1}^{1}\frac{ dz}{2} \frac{v_s k_{F}z+t + (\Delta z+S)\tau_x}{\sqrt{(v_s k_F z  + t )^2-(\Delta z +S)^2}},
\end{equation}
where $\tau = 1/2\pi\nu\gamma$ is the mean free time. 
Using the condition $|v_s|k_F \ll \Delta $, we obtain two equations for $t$ and $S$:
\begin{align}\nonumber
&i 2\tau \Delta =\mathrm{arcsin}\frac{\Delta}{t} \\
&S = -\frac{i}{2\tau \Delta}\left[ t\frac{v_s k_F}{\Delta} \mathrm{arcsin}\frac{\Delta}{t} + \frac{t v_s k_F - \Delta S }{\sqrt{t^2-\Delta^2}}\right],
\end{align}
which together give
\begin{equation}
t= \frac{-i \Delta}{\mathrm{sh}(2\tau\Delta)},~~ S = t \frac{v_s k_F}{\Delta}.
\end{equation}
Hence one recovers the expression in Eq. (\ref{polar_z}) of the main text
\begin{equation}
\Sigma = \frac{-i \Delta}{\mathrm{sh}(2\tau\Delta)} \left(1 + \tau_x \frac{v_s k_F}{\Delta}\right).
\end{equation}

%%%%%%%%%%%%%%%%%%%%%%%%%%%%%%%%%%%%%%%%%%%%%%%%%%%%%%%%%%%%%%%%%%%%%
%%%%%%%%%%%%%%%%%%%%%%%%%%%%%%%%%%%%%%%%%%%%%%%%%%%%%%%%%%%%%%%%%%%%%
%%%% Bibliography
%\bibliographystyle{apsrev4-1}
\bibliography{NonHermitianSC_PRB_resub}
%\bibliography{NonHermitianSC}
%%%%%%%%%%%%%%%%%%%%%%%%%%%%%%%%%%%%%%%%%%%%%%%%%%%%%%%%%%%%%%%%%%%%%
%%%%%%%%%%%%%%%%%%%%%%%%%%%%%%%%%%%%%%%%%%%%%%%%%%%%%%%%%%%%%%%%%%%%%

\end{document}